\documentclass[12pt]{iopart}
\usepackage{graphicx}
\usepackage{booktabs}
\usepackage{hyperref}
\usepackage{xspace}
\usepackage{amssymb}

\begin{document}

\title{Understanding the core density profile in TCV H-mode plasmas}

\author{
D. W\'agner,
E. Fable$^1$,
A. Pitzschke,
O. Sauter,
H. Weisen
and the TCV team}

\address{
Ecole Polytechnique F\'ed\'erale de Lausanne (EPFL),
Centre de Recherches en Physique des Plasmas,
Association Euratom-Conf\'ed\'eration Suisse, Station 13, CH-1015 Lausanne,
Switzerland \\

$^1$Max-Planck-Institut f\"ur Plasmaphysik, IPP-EURATOM Association
Boltzmannstra\ss e 2
D-85748 Garching bei M\"unchen, Germany
}

\ead{david.wagner@epfl.ch}

\begin{abstract}
Results from a database analysis of H-mode electron density profiles on the
Tokamak \`a Configuration Variable (TCV) under stationary conditions show
that the logarithmic electron density gradient increases with
collisionality.  By contrast, usual observations of H-modes showed that the
electron density profiles tend to flatten with increasing collisionality.
In this work it is reinforced that the role of collisionality alone,
depending on the parameter regime, can be rather weak and in these,
dominantly electron heated TCV cases, the electron density gradient is
tailored by the underlying turbulence regime, which is mostly determined by
the ratio of the electron to ion temperature and that of their gradients.
Additionally, mostly in ohmic plasmas, the Ware-pinch can significantly
contribute to the density peaking.  Qualitative agreement between the
predicted density peaking by quasi-linear gyrokinetic simulations and the
experimental results is found.  Quantitative comparison would necessitate
ion temperature measurements, which are lacking in the considered
experimental dataset.  However, the simulation results show that it is the
combination of several effects that influences the density peaking in TCV
H-mode plasmas.
\end{abstract}

\newcommand{\rln}{\ensuremath{\frac{R}{L_n}}}
\newcommand{\rlne}{\frac{R}{L_{n_{\mathrm{e}}}}}
\newcommand{\rlTe}{\frac{R}{L_{T_e}}}
\newcommand{\rlTi}{\frac{R}{L_{T_i}}}
\newcommand{\ilrln}{\ensuremath{R/L_n}\xspace}    % inline r/ln
\newcommand{\rlnstat}{\ensuremath{\left. \rln \right|_\mathrm{stat}}} % inline r/ln stat
\newcommand{\ilrlnstat}{\ensuremath{\left. \ilrln \right|_\mathrm{stat}}\xspace} % inline r/ln stat
\newcommand{\ilrlTe}{R/L_{T_e}\xspace}   % inline r/te
\newcommand{\ilrlTi}{R/L_{T_i}\xspace}   % inline r/te
\newcommand{\ilgradTrat}{L_{T_\mathrm{i}}/L_{T_\mathrm{e}}\xspace}

\newcommand{\CP}{C_\mathrm{P}}
\newcommand{\CT}{C_\mathrm{T}}

\newcommand{\Bzero}{B_\mathrm{0}}

\newcommand{\dense}{n_{\mathrm{e}}}
\newcommand{\densi}{n_{\mathrm{i}}}
\newcommand{\Te}{T_{\mathrm{e}}}
\newcommand{\Ti}{T_{\mathrm{i}}}
\newcommand{\TeTi}{\frac{\Te}{\Ti}}
\newcommand{\ilTeTi}{\Te/\Ti}

\newcommand{\GammaGS}{\Gamma^\mathrm{GS2}_\mathrm{e}}
\newcommand{\GammaQL}{\left<\Gamma\right>}
\newcommand{\qe}{Q_\mathrm{e}}
\newcommand{\qeQL}{\left<Q_\mathrm{e}\right>}
\newcommand{\qiQL}{\left<Q_\mathrm{i}\right>}
\newcommand{\qeEXP}{Q^\mathrm{exp}_\mathrm{e}}

\newcommand{\Chie}{\chi_\mathrm{e}}
\newcommand{\ChiePB}{\chi_\mathrm{e}^{\mathrm{PB}}}
\newcommand{\Wp}{W_\mathrm{p}}
\newcommand{\omegaQL}{\left<\omega_\mathrm{r}\right>} 
\newcommand{\omegaR}{\omega_\mathrm{r}} 

\newcommand{\vnewk}{\hat{\nu}}
\newcommand{\nueff}{\nu_\mathrm{eff}}
\newcommand{\Zeff}{Z_\mathrm{eff}}

%thermal speed
\newcommand{\vthi}{v_\mathrm{th}^\mathrm{i}} 
\newcommand{\vthe}{v_\mathrm{th}^\mathrm{e}} 

%Larmor radii
\newcommand{\rhoi}{\rho_{\mathrm{i}}}

% wavenumber
\newcommand{\ky}{\ensuremath{k_y}}
\newcommand{\aky}{\ensuremath{\ky \rhoi}\xspace}

% common units
\newcommand{\keV}{\mathrm{keV}}
\newcommand{\icm}{10^{19}\ \mathrm{m^{-3}}}

\newcommand{\Rmajor}{R}

\newcommand{\Grho}{\left<\left| \nabla \rho_\psi \right| \right>}

\newcommand{\imagewidth}{0.45\textwidth}
\newenvironment{myfigure}%
    {\begin{figure}[p!] \begin{center}}%
    {\end{center}\end{figure}}

\newcommand{\panel}[1]{(\emph{#1})}

\newcommand{\TCVparameters}
{$R=0.88\,\mathrm{m}$, $a=0.25\,\mathrm{m}$, %
$I_\mathrm{p} \leqslant 1\,\mathrm{MA}$,%
$B_\mathrm{t} \leqslant 1.5\,\mathrm{T}$, $\kappa \leqslant 2.8$}

\newcommand{\OHreference}{$\ilTeTi=1.0$, $\vnewk=0.024$}
\newcommand{\ECHreference}{$\ilTeTi=2.0$, $\vnewk=0.008$}

\section{Introduction}

In recent years numerous works have been devoted to particle transport studies
(see the most recent review \cite{angioni2009particle_transport} and references
therein).  The experimental observations and the theoretical predictions show a
good overall agreement.  It is now widely accepted that the electron density
profile in the core of tokamak plasmas is mainly tailored by turbulent
mechanisms.  The turbulent state present in the plasma is intimately related to
the density peaking and its dependence on the plasma parameters.  In this work
we apply a unified quasilinear theory of ion temperature gradient (ITG) and
trapped electron (TEM) microinstabilities and turbulence \cite{fable2010role}
that has emerged in the last years to a very specific problem, namely the
density profile behaviour in TCV core plasmas.  The success of quasi-linear
theory in explaining particle transport in TCV L-mode and eITB plasmas
\cite{fable2010role, fable2008eITB} motivates us to test the theoretical
predictions against an H-mode dataset as well, using the same methodology.  Our
goal is to reproduce the observed experimental trends with numerical simulations
and understand which transport mechanisms are influential on the density
peaking.

Alcator C-Mod internal transport barriers (ITBs) with very strongly peaked
density profiles showed density profile flattening with modest electron
heating, which was explained by the onset of TEM turbulence, using linear
and nonlinear gyrokinetic simulations \cite{ernst2004role}.  The density
peaking in the C-Mod ITBs was thought to result from the Ware pinch, with
the ITB density gradient during electron heating determined by the balance
of Ware pinch and TEM turbulent fluxes, thus acquiring a strong inverse
dependence on temperature.  In TCV electron ITBs, where transport barriers
are seen mainly on the electron temperature profile \cite{coda2007physics},
the density peaking results from a strong thermodiffusive pinch
\cite{fable2006inward} component which can be explained \cite{fable2008eITB}
using the quasi-linear model that shall be used in this paper .

The first experiments on TCV with third-harmonic X-mode electron cyclotron
resonance heating (X3 ECH) \cite{alberti2005third, porte2007plasma} reported
reduction of peakedness of the density profile when intense electron heating
was applied.  A preliminary study \cite{maslov2006eps} of these plasmas
concluded that the density peaking is mainly caused by turbulent processes
and suggested that the intense electron heating being favourable for TEM
destabilization can possibly lead to profile flattening.  In recent
experimental campaigns of TCV, new data have been collected on ohmic (OH)
and ECH H-modes, repeating the very first experiments and also exploring a
wide parameter range.  These experiments, where special attention was made
to the quality of the collected data, allow us to better explore the
experimental dependencies and to have more confidence in validation of the
theory.  The experimental dataset is used to test the quasi-linear model
proposed in \cite{fable2010role}.  Our work is complementary to the recent
study with AUG data \cite{angioni2011hmodes}.

In the next section a brief overview of the TCV H-mode experiments is given
with a particular attention to the typical parameter ranges within which
these plasmas are sustained.  In section \ref{sec:theoretical-model} the
theoretical model that is used for the interpretation of the experiments is
summarized.  In section \ref{sec:simulation_results} the results of the
simulations are discussed and compared to the experimental observations.
Finally, in section \ref{sec:discussion}, general discussion and concluding
remarks are given.

\section{TCV H-mode plasmas}
\label{sec:experiments}

In TCV (\TCVparameters), L-mode to H-mode transitions are obtained with and
without additional heating \cite{hofmann1994creation,
martin2003accessibility}.  The additional heating is provided by powerful
third-harmonic electron cyclotron heating system (X3 ECH, 118\,GHz,
$3\times0.5\,\mathrm{MW}$, 2\,s, top launch) \cite{alberti2005third}.  In
this work we focus on standard ELMy H-modes with type I and type III ELMs
with typical ELM frequencies around 35--100\,Hz.  Time traces of a
typical pulse are shown in figure \ref{fig:example-pulse}.  More `exotic'
scenarios such as the quiescent ELM-free \cite{porte2007plasma} and
snowflake \cite{piras2010snowflake} H-mode scenarios are beyond the scope of
this present paper and not considered in the analysis.

The experimental database is built on a representative set of sufficiently
diagnosed H-mode deuterium pulses.  Some typical parameters of these plasmas in
the database are shown in Table \ref{tab:typical-parameters}.  The dataset
contains pulses with OH only and those with additional X3 ECH, the different
power levels are relatively well covered.  The dataset is somewhat less extended
in terms of changes in plasma current and contains plasmas with very similar
shape.

% Table: typical plasma parameters
\newcommand{\trow}[3]{% \trow{quantity, lo, hi}
#2 &$<$& #1 &$<$& #3\\}

\begin{table}
    \centering
    \begin{tabular}{ccccc}
    \toprule
    \trow{$\left<\dense \right>_\mathrm{vol} [10^{19} \mathrm{m^{-3}}]$}{3.7}{6.1}
    \trow{${\Te}(0) [\mathrm{keV}]$}{0.9}{2.5}
    \trow{$I_\mathrm{p}\ [\mathrm{kA}]$}{280}{420}
    \trow{$V\ [\mathrm{m^3}]$}{1.35}{1.55}
    \trow{$P_\mathrm{OH}\ [\mathrm{kW}]$}{150}{600}
    \trow{$P^\mathrm{inj}_\mathrm{ECH}\ [\mathrm{kW}]$}{250}{1000}
    &\multicolumn{3}{c}{%
        $\delta_\mathrm{edge}\approx0.45$, $\kappa_\mathrm{edge}\approx1.7$, %
        $q_{95}\approx 2.4$%
    } &\\
    \bottomrule
    \end{tabular}
    \caption{Typical parameters of the TCV H-modes.}
    \label{tab:typical-parameters}
\end{table}

\begin{myfigure}
    \includegraphics[width=\imagewidth]{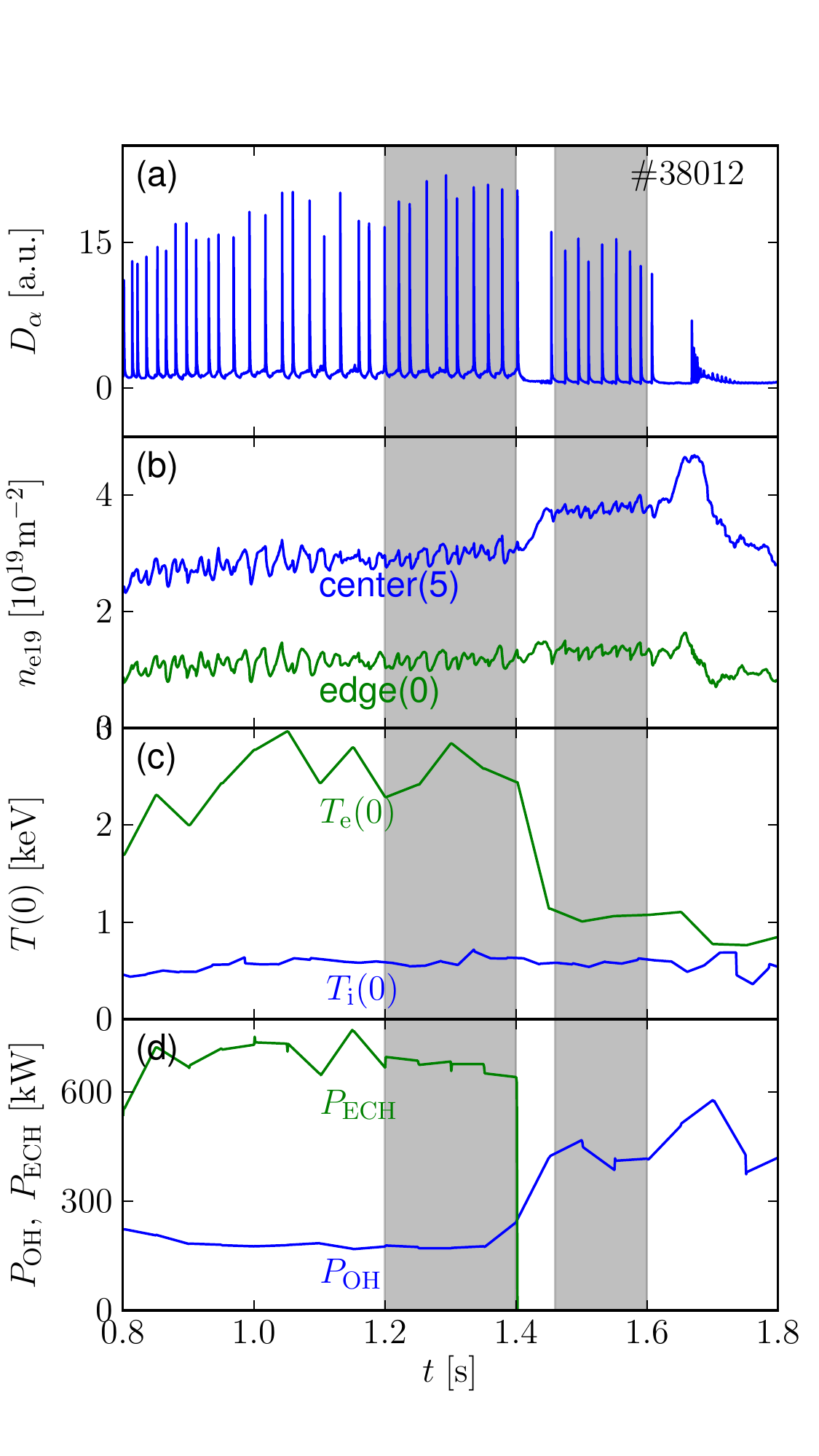}
    \caption{
    Time traces of
    \panel{a} D$_\alpha$ emission;
    \panel{b} central and peripheral FIR chords;
    \panel{c} central electron and ion (C$^{6+}$) temperature;
    \panel{d} OH and ECH power.  The stationary profiles
    are taken from the shaded intervals \mbox{1.2--1.4 s} and \mbox{1.46--1.6
    s}.
    }
    \label{fig:example-pulse}
\end{myfigure}

The majority ion species is deuterium and the most abundant impurity is
carbon, originating from the wall tiles.  The typical carbon concentration
is a few per cent of the electron density. The effective charge of these
plasmas, estimated from soft X-ray measurements
\cite{weisen1991measurement}, is found to be $\Zeff \approx 2$ for all the
pulses considered in the analysis, independently on the heating scheme
applied.  The calculated total (OH plus bootstrap) current, assuming
neoclassical conductivity with the measured $\Zeff$ value, is consistent
with the magnetic measurements.

\subsection{Density and electron temperature measurements}
\label{ssec:density-temperature-measurements}

The electron density and temperature profiles are measured via Thomson
scattering.  Density profile measurements are cross-calibrated with the far
infrared interferometer (FIR).  Typical temperature and density profiles are
depicted in figure \ref{fig:profiles}.  The radial flux label throughout this
paper is
$\rho_\psi = \sqrt{(\psi - \psi_0)/(\psi_\mathrm{b} - \psi_0)}$,
where $\psi$ is the poloidal flux, $\psi_0$ and $\psi_\mathrm{b}$ are the
poloidal flux at the magnetic axis and the plasma boundary, respectively. The
derivatives of the profiles with respect to $\rho_\psi$ are determined from
cubic spline fits.  The fit was performed on profiles collected during a
stationary phase of pulses, i.e.~where the main plasma parameters (plasma
current, line averaged density, internal inductance) do not change
significantly.  The normalized inverse scale length (normalized gradient) of a
profile is defined as:
$R/L_X = - R \Grho \partial \log{X}/\partial \rho_\psi$,
where $R$ is the major radius and the $\Grho$ is the Jacobian of the
$r\rightarrow \rho_\psi$ coordinate transformation with $r$ being the distance
from the magnetic axis on the outboard midplane.  The profiles of the
normalized gradients are shown in figure \ref{fig:profile_gradients}.

All the considered pulses exhibited sawteeth.  This results in flat average
profiles within $\rho_\psi \approx 0.4-0.6$ surface, depending on $I_\mathrm{p}$.

\subsection{Ion temperature measurements}
\label{ssec:ion_measurements}

The $\mathrm{C}^{6+}$ ion temperature is measured by charge-exchange
recombination spectroscopy (CXRS) \cite{bortolon2009}.  Neutral particle
analysers (NPA) \cite{karpushov2006neutral} provide estimate of the central
ion temperature.  In these H-mode plasmas, however, due to geometrical
constraints, the ion temperature measurements are limited outside $\rho_\psi
\approx 0.8$.  This fact, strictly speaking, does not allow us to measure
the ion temperature profile in the core.

figure \ref{fig:profiles} b) also shows a typical ion temperature profile between
$\rho_\psi = 0.8 - 1$ overlaid on the electron temperature profiles.  The ion
temperature does not change significantly when the ECH is applied.  From these
typical profiles the local value of $\ilTeTi$ at $\rho_\psi = 0.7$
($\rho_\mathrm{vol}\approx 0.6$) is around $1$
for phases with ohmic heating only, and about $2$ for phases with ECH.

The present ion temperature data suggests that the ion temperature gradient
is lower than that of the electrons.  Assuming the functional form
$\Ti(\hat{\rho})\left(\Te/\Te(\hat{\rho}) \right)^{\nu}$ for the ion
temperature profile, the relation $\ilrlTi = \nu \ilrlTe$ follows.  In
figure \ref{fig:profiles} b) the dashed curve, derived from the electron
temperature profile of \#40089 using $\hat{\rho}=0.8$ and $\nu=0.5$, follows
rather well the ion temperature measurements outside $\rho_\psi \approx
0.8$.  Following these considerations we estimate $\ilrlTi$ to be around
3--5 at $\rho_\psi=0.7$.

\begin{myfigure}
    \includegraphics[width=\imagewidth]{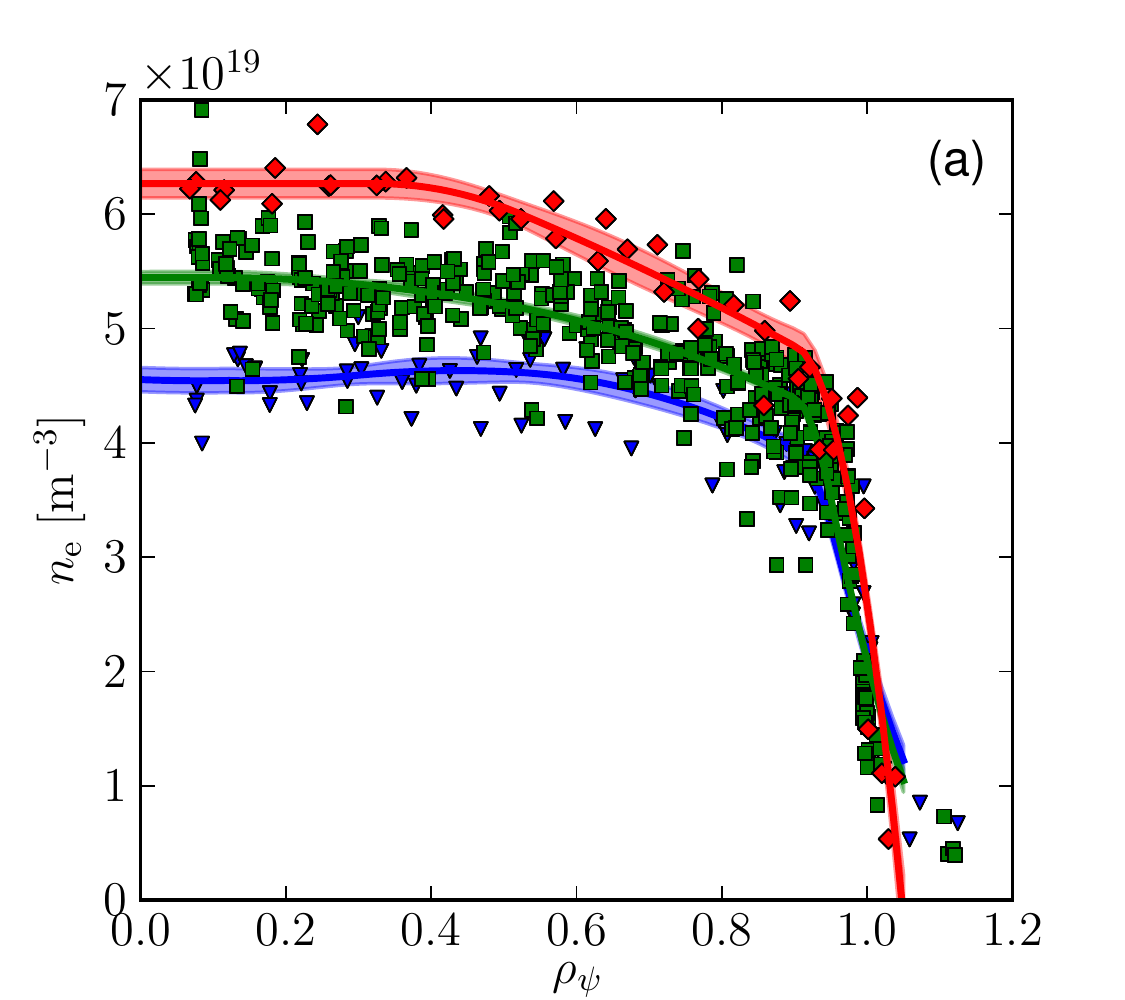}
    \includegraphics[width=\imagewidth]{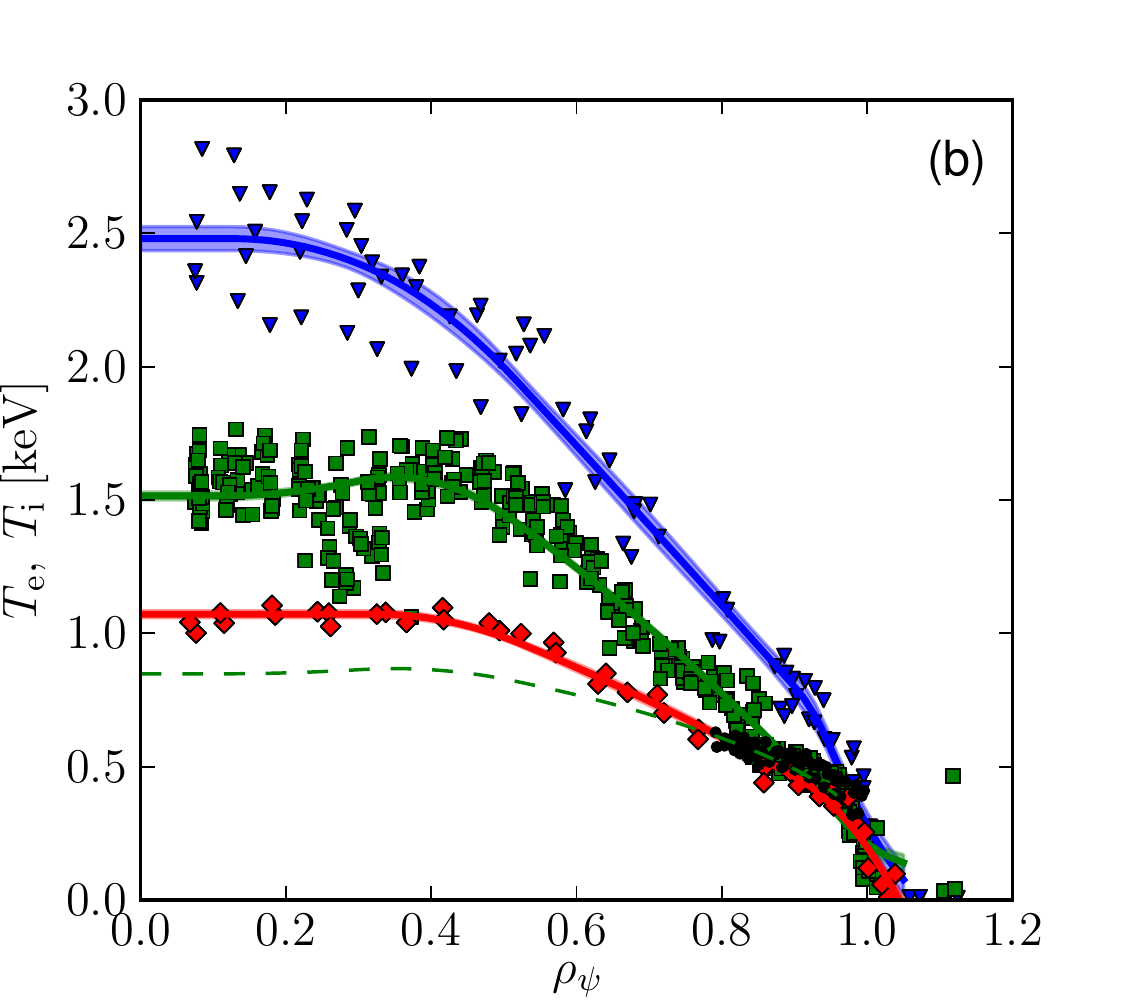}
    \caption{
    Profiles comparing conditions with ohmic heating only (diamonds,
    \mbox{\#38012 $1.46-1.6\ \mathrm{s}$} ) to those with
    additional
    $350\ \mathrm{kW}$ (squares, \mbox{\#40089, $0.7 - 1.61\ \mathrm{s}$}),
    $675\ \mathrm{kW}$ (triangles pointing down, \mbox{\#38012 $1.2-1.4\
    \mathrm{s}$})
    ECH.
    \panel{a} electron density,
    \panel{b} electron temperature, ion temperature
        (points, \mbox{\#40089, $0.7 - 1.61\ \mathrm{s}$}).
    Solid lines show the piecewise polynomial fits
    which were used to compute the gradients.  The shaded areas mark the
    confidence intervals corresponding to 95\% confidence level.  The dashed
    line represents $\Ti(0.8)\left(\Te/\Te(0.8) \right)^{0.5}$
        (\mbox{\#40089, $0.7 - 1.61\ \mathrm{s}$}).
    }
    \label{fig:profiles}
\end{myfigure}

\begin{myfigure}
    \includegraphics[width=\imagewidth]{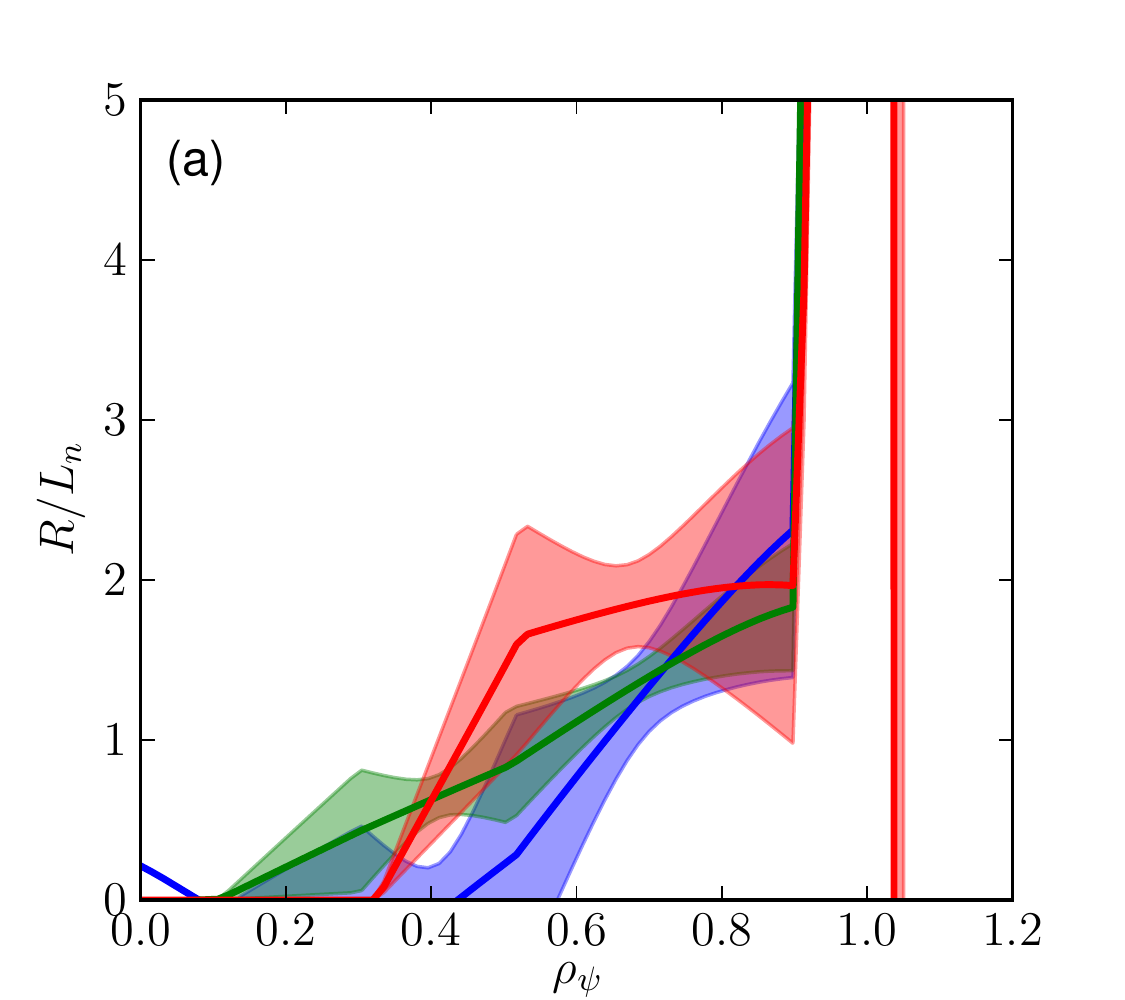}
    \includegraphics[width=\imagewidth]{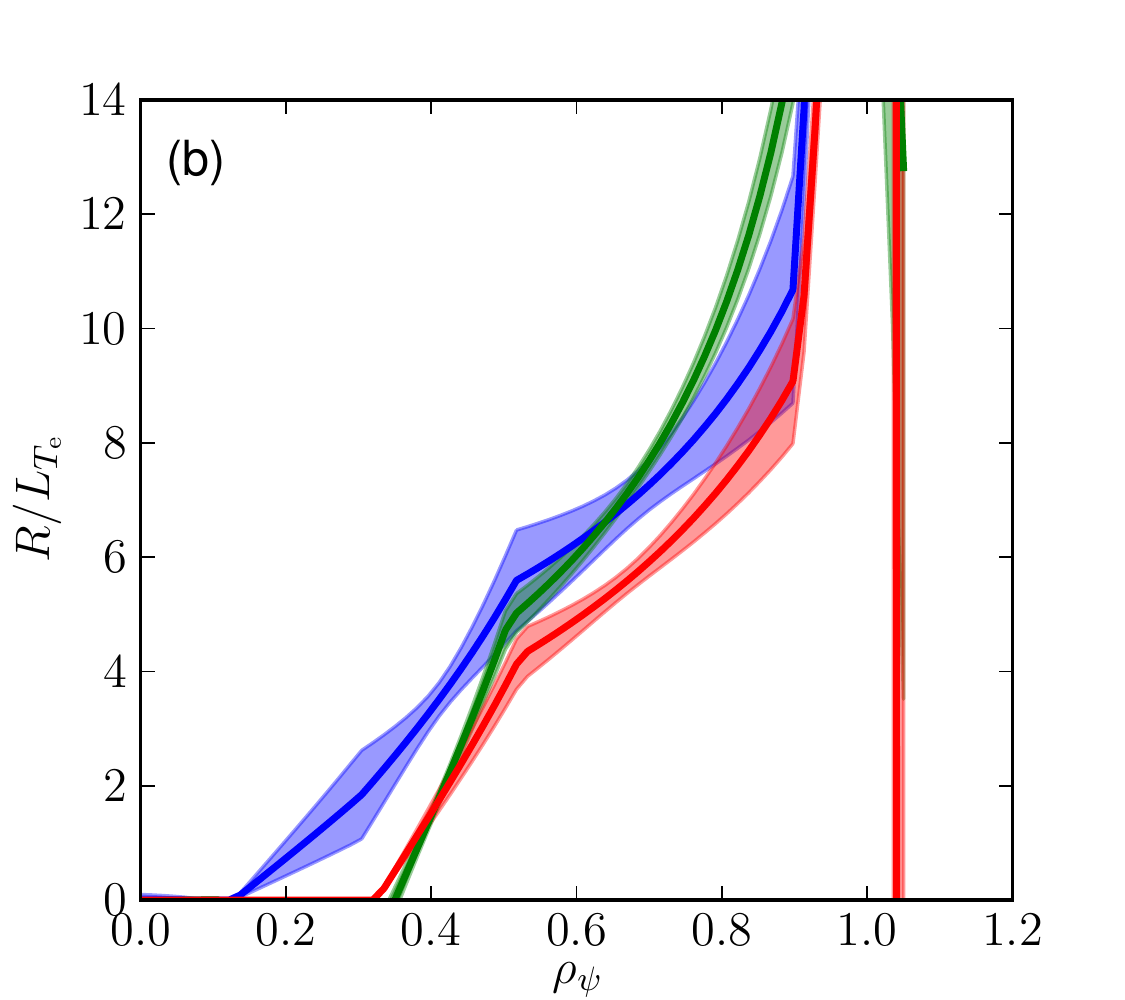}
    \caption{
    Profiles of the normalized logarithmic gradients calculated from the
    piecewise polynomial fits depicted in figure \ref{fig:profiles}.  The shaded
    areas mark the confidence intervals corresponding to 95\% confidence level.
    }
    \label{fig:profile_gradients}
\end{myfigure}

\subsection{Collisionality dependence}
\label{ssec:collisionality-dependence}

Statistical analysis of JET \cite{weisen2006scaling} and AUG
\cite{angioni2007scaling} showed that collisionality is the most important
scaling parameter for density peaking in H-modes.  Both machines report the
flattening of the density profile with increasing collisionality.  Note that in
these machines collisionality is controlled by the combination of neutral beam
injection (NBI) heating and ECH.  Despite its complexity, collisionality is a
frequently used parameter to map parameter dependencies and for inter-machine
comparisons.  In this subsection the collisionality dependence of the density
and temperature profile peaking is presented.

Later (in section \ref{sec:simulation_results}), the results of the numerical
modelling will be presented as a function of the following definition of
collisionality:
\begin{equation}
    \vnewk = \nu_{\mathrm{ei}}/(v_\mathrm{th}/R),
    \label{eq:collisionality}
\end{equation}
where $\nu_{\mathrm{ei}}$ is the electron-ion collision frequency,
$v_\mathrm{th}=\sqrt{\Ti/m}$ the thermal velocity and $\Rmajor$ the major radius.
In order to facilitate the comparison with previous results the experimental
dependences are also plotted as a function of an effective collisionality
adopted from \cite{angioni2005relationship}:
$\nueff = 0.1 \dense \Zeff \Rmajor/\Te^2
    (\approx\nu_\mathrm{ei}/\omega_\mathrm{De})$,
where $\dense$, $\Te$ are the electron density in $\icm$ and the electron
temperature in $\keV$ with $\Zeff=2$.

figure \ref{fig:peakedness_vs_vnewk} shows the normalized logarithmic
gradients of the electron temperature ($\ilrlTe$) and density ($\ilrln$)
profiles at $\rho_\psi = 0.7$ as a function of collisionality $\vnewk$ and
$\nueff$.  Points around $\vnewk\approx 0.02$, the most typical TCV OH
H-mode pulses, have $\ilrln$ around 3.  Samples from reversed field
campaigns with unfavourable ion $\nabla B$ drift direction with somewhat
higher density, are located at around $\vnewk=0.03$ having $\ilrln \approx
3.5$.  In the $0.008 < \vnewk < 0.02$ range, which is reached by adding
0.5-1.5 MW ECH, $\ilrln$ decreases to about $1.5$.

Note that the global trend of $\ilrln$ as a function of the collisionality is
very similar to the one observed on FTU fully non-inductive, high density
electron heated plasmas \cite{romanelli2007parametric}.  By contrast, H-modes at
JET \cite{weisen2006scaling} and AUG \cite{angioni2011hmodes} showed that
$\ilrln$ is decreasing with increasing collisionality.

\begin{myfigure}
    \includegraphics[width=\imagewidth]{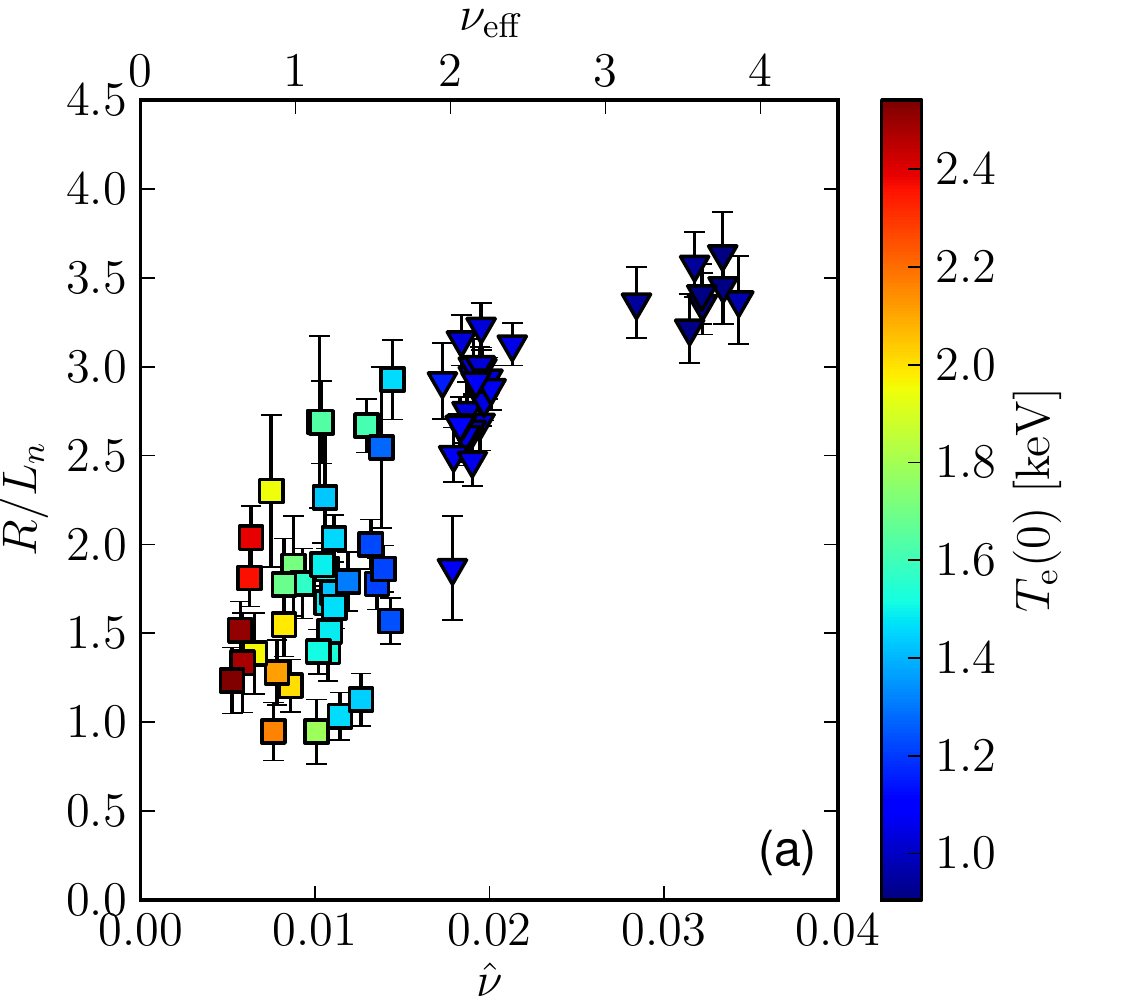}
    \includegraphics[width=\imagewidth]{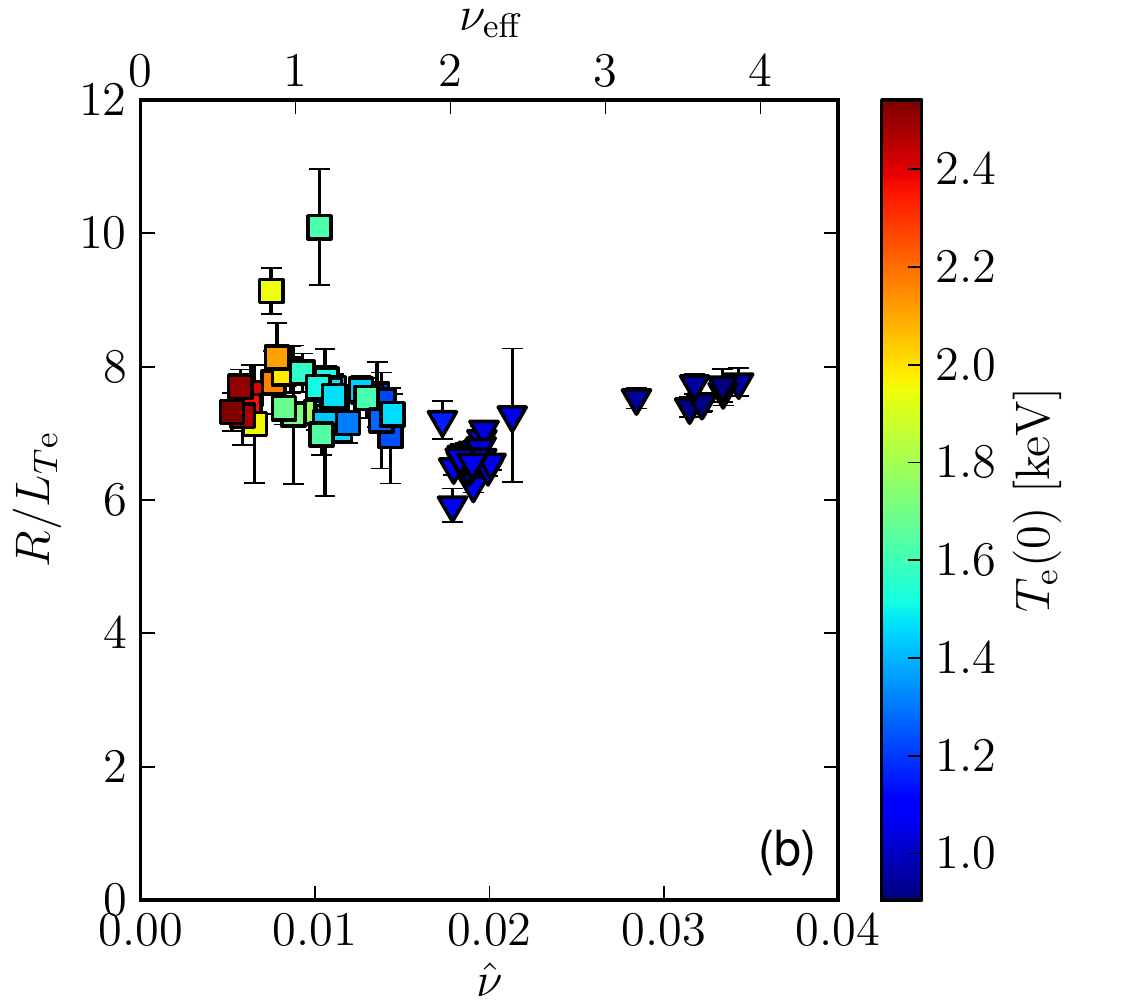}
    \caption{
    Scatter plots of the normalized logarithmic gradients at $\rho_\psi=0.7$ as
    a function of collisionality over the considered database.  \panel{a}
    $\ilrln$ \panel{b} $\ilrlTe$.  Different symbols indicate different heating
    phases, with ohmic heating only (triangles pointing down) and  with central
    X3 ECH (squares).  The colorscale (online) represents $T_\mathrm{e}(0)$, the
    central value of the electron temperature.
    }
\label{fig:peakedness_vs_vnewk}
\end{myfigure}

\section{Theoretical model}
\label{sec:theoretical-model}

In Ref.~\cite{fable2010role} a quasi-linear gyrokinetic framework has been
proposed for particle transport with which TCV L-mode plasmas and electron
internal transport barriers \cite{fable2008eITB} were studied extensively.
In this paper we strive to test this model against TCV H-mode data to
understand the density profile behaviour.  Although there is no general
agreement in the community on the applicability of different quasi-linear
models, significant effort has been made to compare various models against
non-linear simulations \cite{casati2009validating, merz2010nonlinear,
bourdelle2007new, lapillonne2011nonlinear} and good agreement was
found.  It was also shown that the fully developed turbulence preserves many
features of the linear evolution \cite{merz2010nonlinear}, therefore with an
ad-hoc weighting of the linear spectra, non-linear fluxes can be acceptably
reproduced \cite{bourdelle2007new}.  Moreover, quasi-linear predictions fit
very well to experimental observations \cite{fable2010role, fable2008eITB,
angioni2011hmodes}.  A relevant exception to this occurs near marginal
stability, where a nonlinear upshift of the TEM critical density gradient
has been found \cite{ernst2004role}, which increases with collisionality
\cite{ernst2006identification}.  This effect is clearly outside the
framework of quasilinear theory.  EC heated experiments are in general not near
marginal stability to temperature gradient driven TEMs \cite{ernst2009role}
with $\ilrlTe\approx8$ \cite{ryter2001experimental} but may be near marginal
stability to longer wavelength density gradient driven TEMs
\cite{ernst2004role}.  In our dataset, \ilrln ranges from about 1 to
approximately 3.5, with the upper value lying above the density gradient
driven TEM treshold (as shown later in figure \ref{fig:spectra-examples}).

We solve the linear gyrokinetic (GK) equation for electrostatic perturbations
with a wavenumber $\mathbf{k}=(0,k_y)$ to obtain $\omegaR^k$, $\gamma^k$,
$\Gamma^k$, the real and imaginary part of the mode frequency and the particle
flux, respectively, of the most unstable mode.  $\omegaR^k>0$ represents a mode
turning in the ion diamagnetic (ITG) direction, while $\omegaR^k<0$ corresponds
to the electron diamagnetic direction (TEM).

$\omegaR^k$, $\gamma^k$, $\Gamma^k$ are evaluated on a range of $k_y$ .  We then
obtain the quasi-linear fluxes and the average mode frequency of the turbulent
state by a weighted average over the mode spectrum:
\begin{equation}
    \left<R\right> = \left.\int_k{w_k R^k \mathrm{d} k}\right/\int_k{w_k \mathrm{d} k},
    \label{eq:qlrule}
\end{equation}
where $R$ can stand for $\Gamma^k$, $q^k$, $\omega^k$.  The $w_k$ weights are
usually chosen according to a quasi-linear rule of the form:
\begin{equation}
    w_k = A_0 \left( \frac{\gamma}{\left< k_\perp^2 \right>}\right)^\xi.
    \label{eq:power_weights}
\end{equation}
We use the same parameters $\xi=2$, $A_0=1$ as in \cite{fable2010role}.  We
shall discuss effect of this choice later, in section \ref{sec:qlrule}.

In order to separate the different mechanisms that are responsible for the
density peaking, one can decompose the particle flux such as:
\begin{equation}
    \Gamma = A \rln + B \rlTe + C,
\end{equation}
which, assuming stationary plasma and using the zero flux condition when no
particle sources are present (as is the case in the core of these ECH/OH
plasmas), transforms into
\begin{equation}
    \rlnstat = -\CT \rlTe - \CP.
    \label{eq:rlnstat}
\end{equation}
The thermodiffusion coefficient $\CT=-B/A$ and the other pinch coefficient
$\CP=-C/A$ are
evaluated numerically \cite{fable2008eITB, fable2010role}.

\section{Simulation results}
\label{sec:simulation_results}

\subsection{Setup for the simulations}
\label{ssec:simulation-setup}

Linear gyrokinetic calculations have been performed with the initial value
flux-tube code GS2 \cite{kotschenreuther1995comparison}.  We introduce two model
cases representing TCV OH and ECH H-mode pulses.  For both $\ilrlTe=9$,
$\ilrlTi=6$, $\Zeff=2$.  For the OH reference case \OHreference, for the ECH
case \ECHreference\ (c.~f.~subsection
\ref{ssec:density-temperature-measurements}).

The magnetic equilibrium of a representative discharge (see
figure \ref{fig:example-pulse}) was calculated by the equilibrium code CHEASE
\cite{lutjens1996chease}.  The $\rho_\psi=0.7$ flux surface with $q=1.2$,
$s=0.7$ was used for flux-tube simulations.  For the collisions both pitch angle
scattering and energy diffusion are taken into account.  13 $k_y$ values are
used in the range of $[0.08, 1.6]$ distributed in a logarithmic way.  All growth
rates and frequencies are normalized to $\vthi/R$.  The binormal wavenumber
$\aky$ is normalized to $\rhoi$, the normal wave number is always zero
($k_x=0$).  Note that on the outboard midplane the normal direction is radial.
The timestep was fixed to $\Delta t = 0.05$ in units of $R/\vthi$.  32 grid
points for each $2 \pi$ turn in $\theta$, and 12 poloidal periods were used.
The parameters are then scanned according to the various simulation results
presented in the next subsections.

\subsection{Finding the stationary density gradient}

\newcommand{\iwf}{.4\textwidth}
\begin{myfigure}
    \includegraphics[width=\iwf]{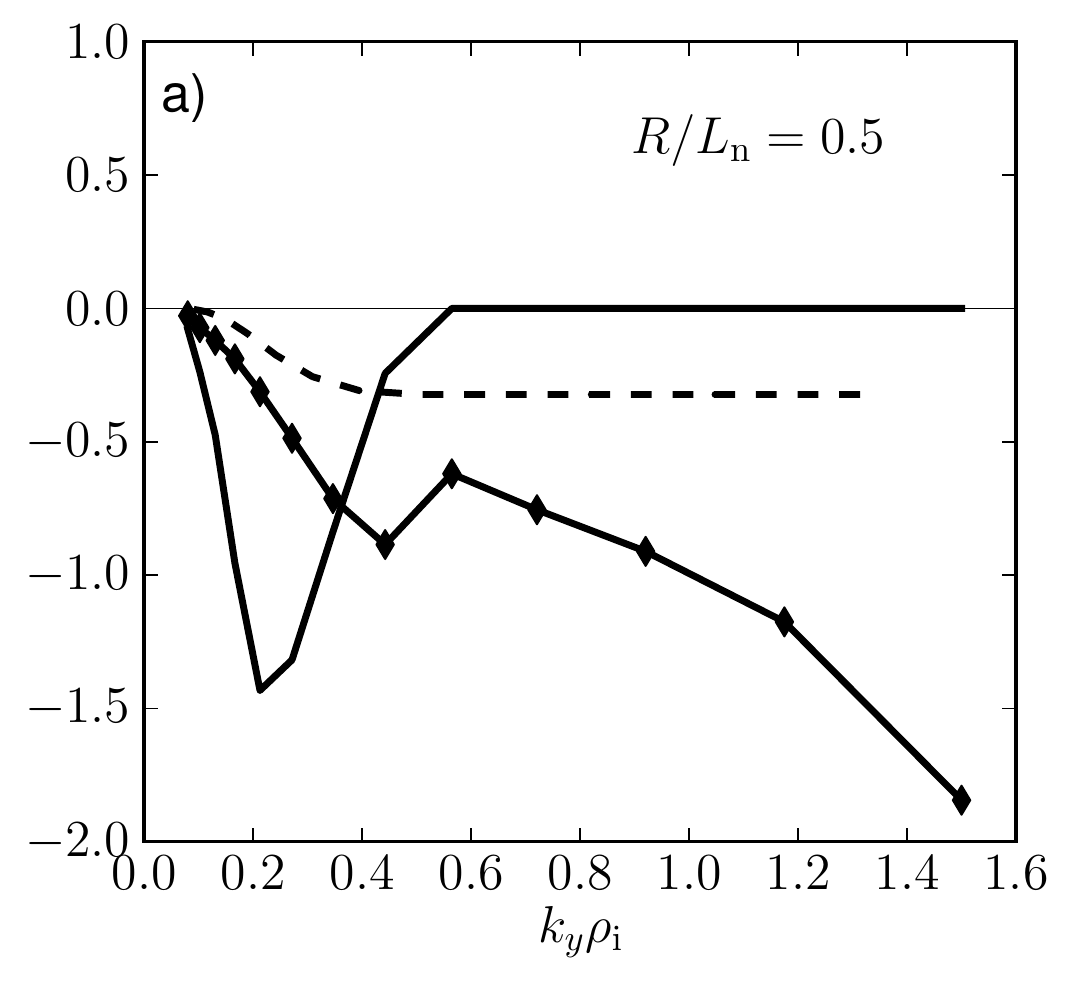}
    \includegraphics[width=\iwf]{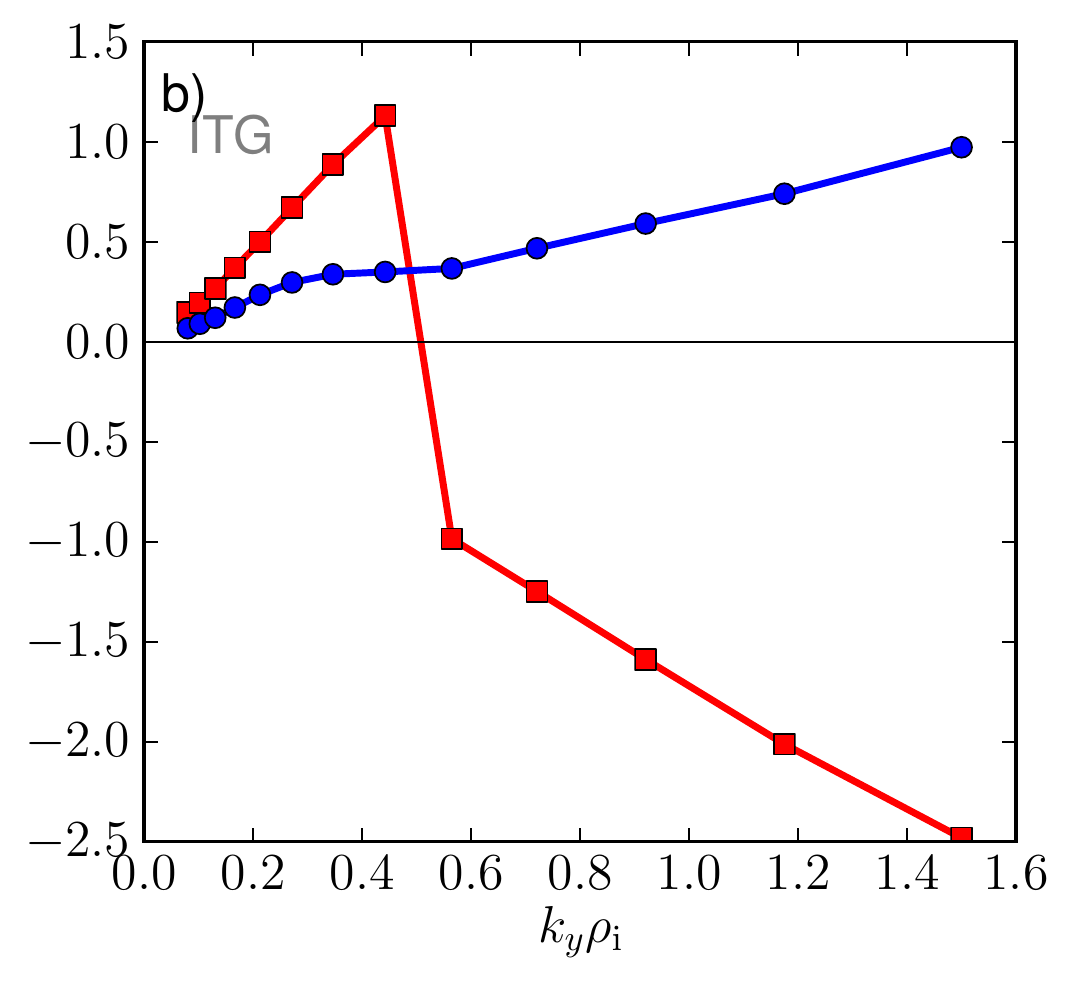}
    \includegraphics[width=\iwf]{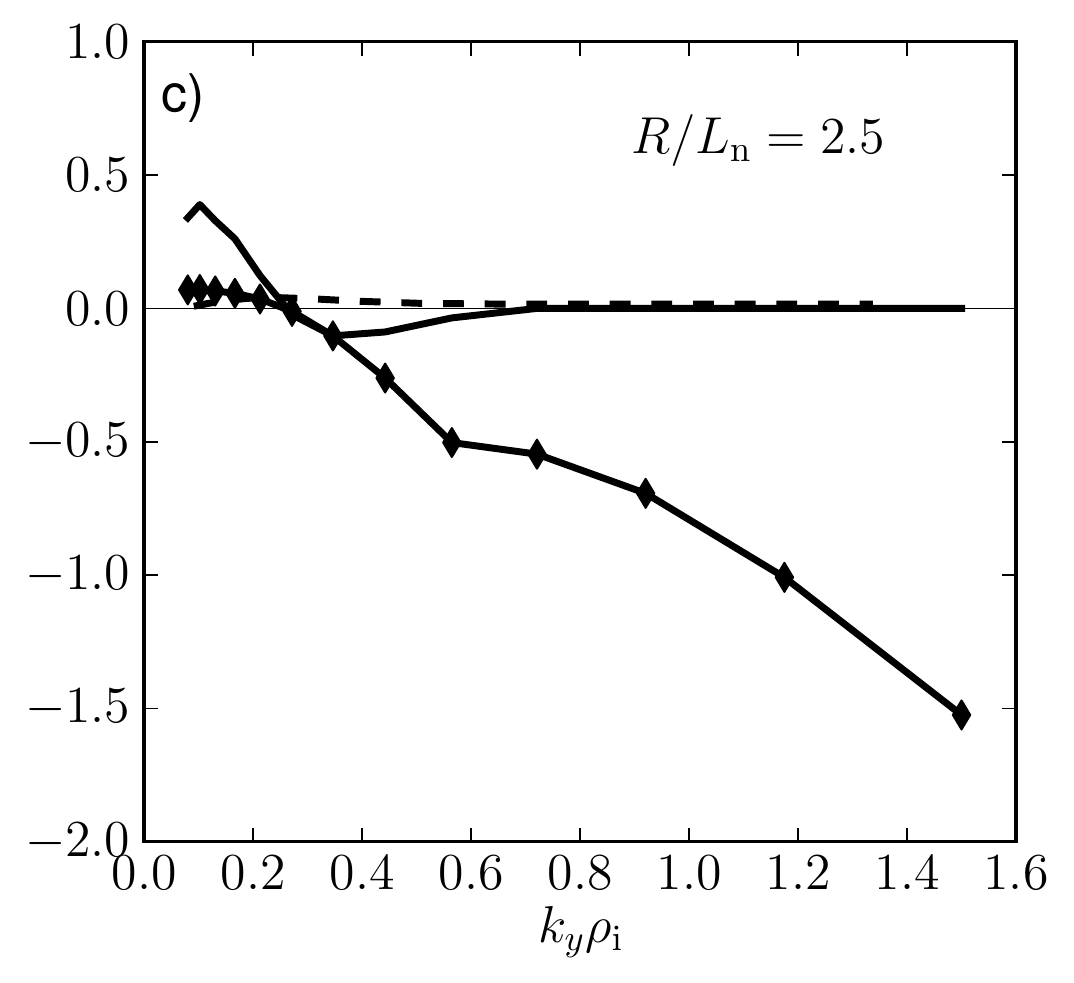}
    \includegraphics[width=\iwf]{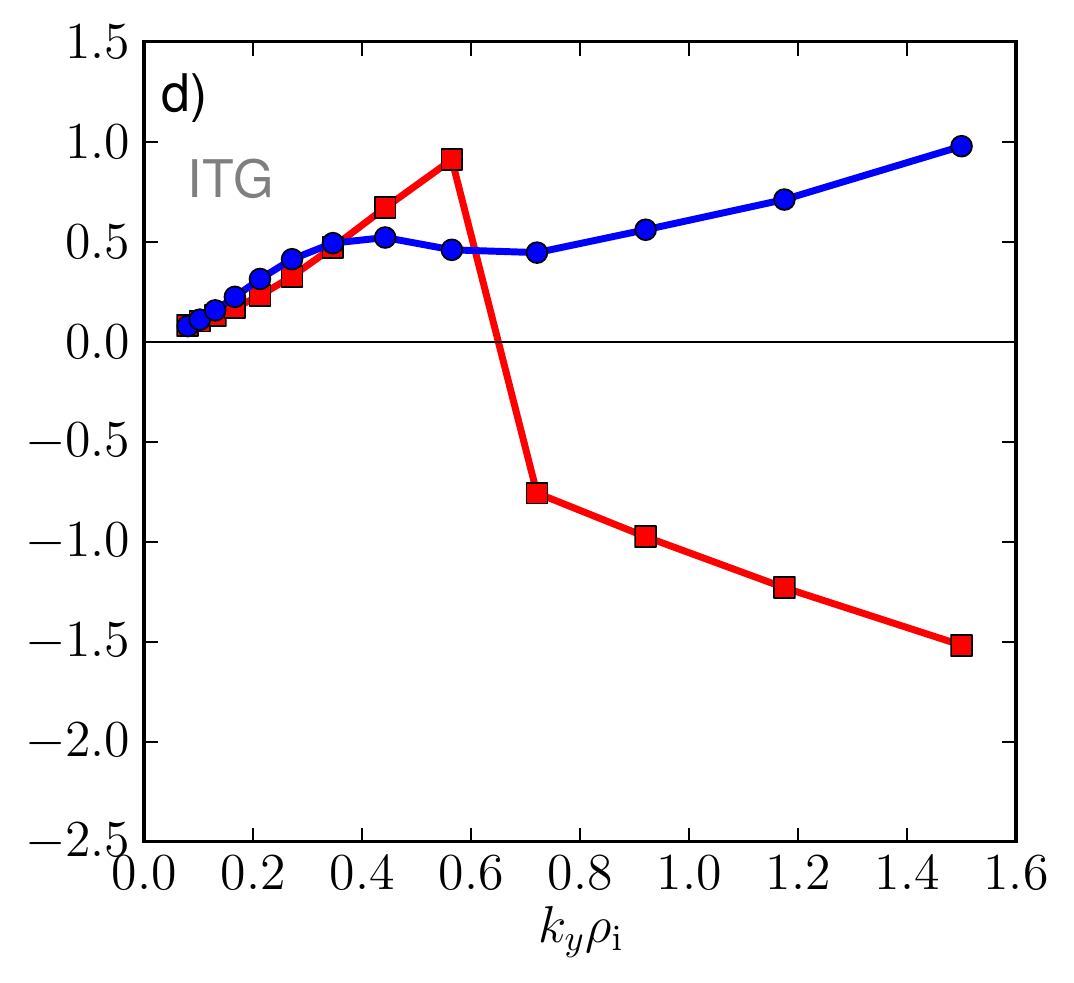}
    \includegraphics[width=\iwf]{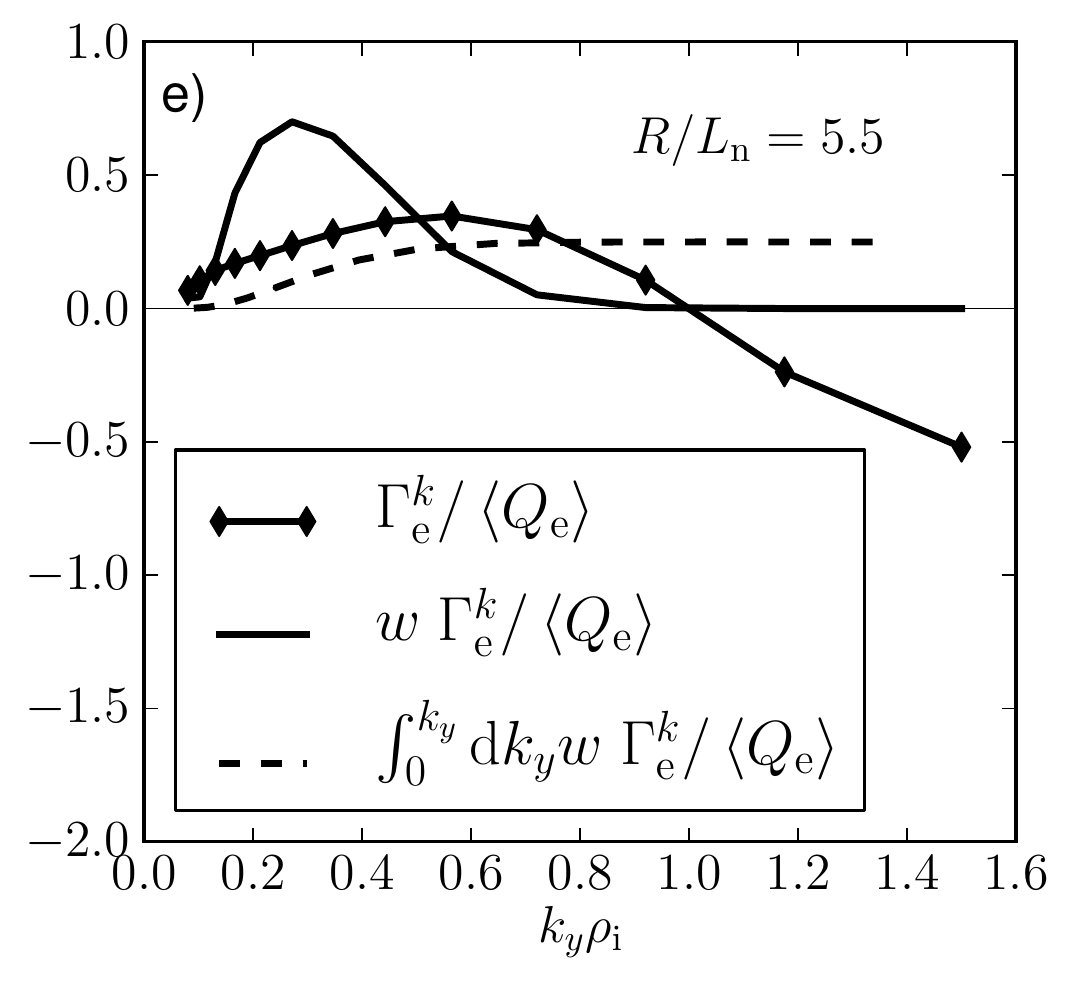}
    \includegraphics[width=\iwf]{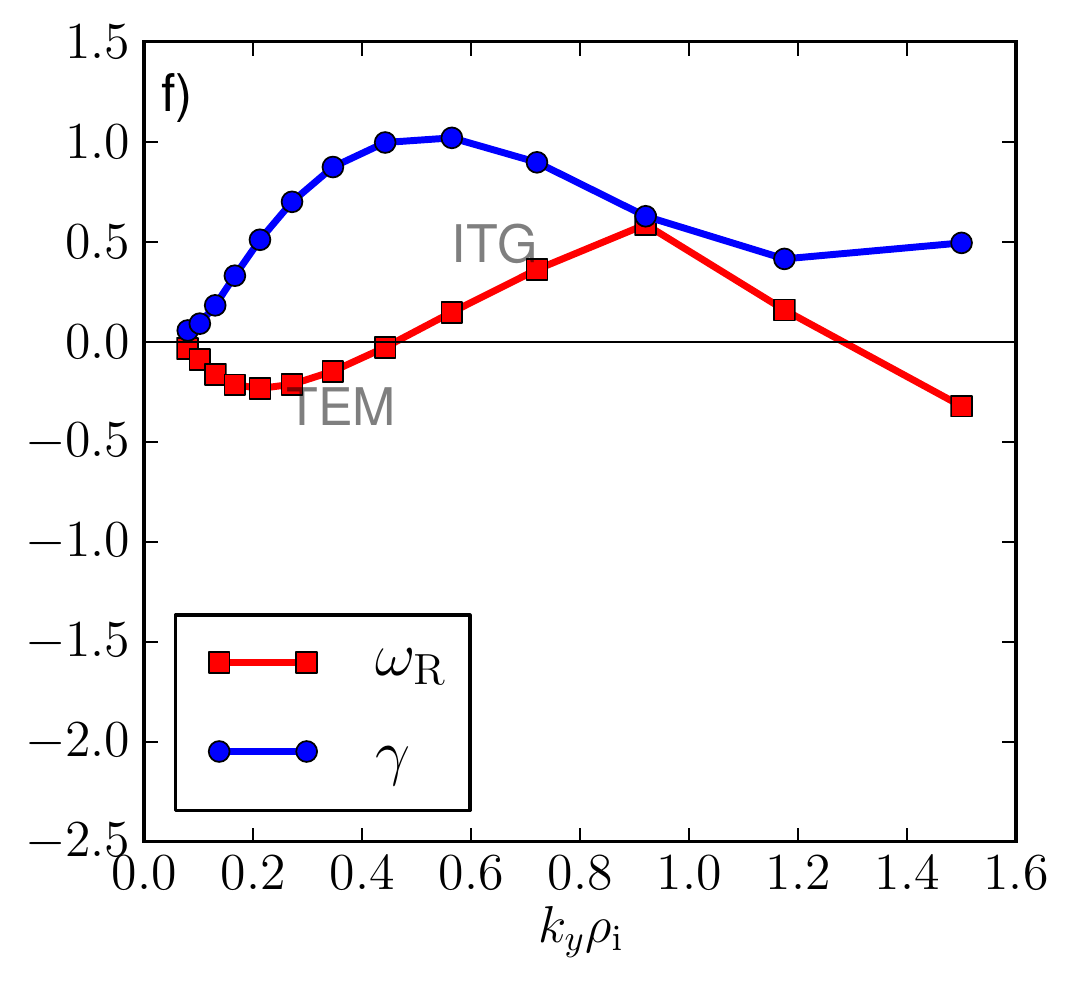}
    \caption{
    Left: particle flux $\Gamma$ (diamonds); right: mode frequency $\omega$ and
    growth rate $\gamma$ (squares and circles, resp.) as a function of
    $\aky$. (a), (b) low $\ilrln$ where ITG is dominant and $\Gamma$
    negative (inwards); (c),(d) $\ilrln \approx \ilrlnstat$ where $\Gamma \approx
    0$; (e), (f) high $\ilrln$ with TEM dominant and positive (outwards)
    particle flux.  The simulation parameters correspond to the OH reference
    case (see section \ref{ssec:simulation-setup}).
    }
    \label{fig:spectra-examples}
\end{myfigure}

\begin{myfigure}
    \includegraphics[width=\imagewidth]{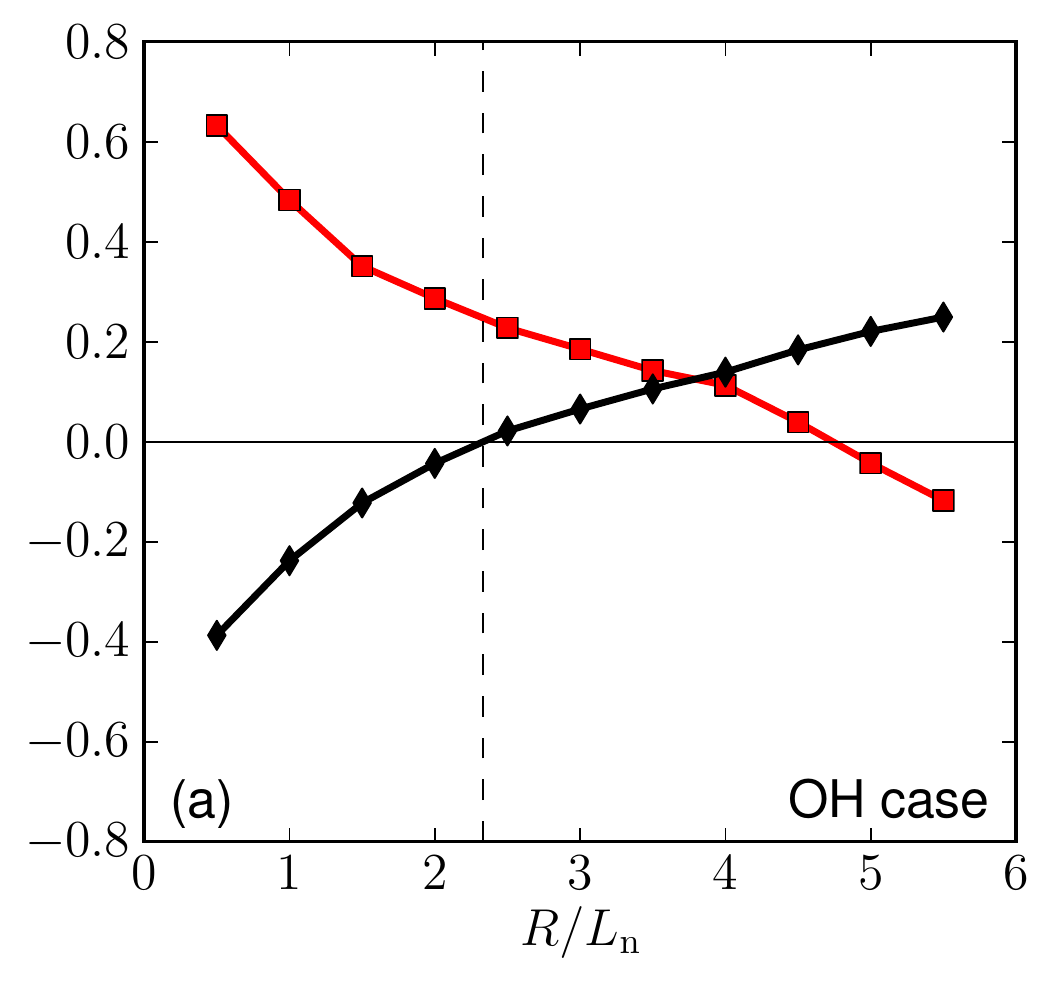}
    \includegraphics[width=\imagewidth]{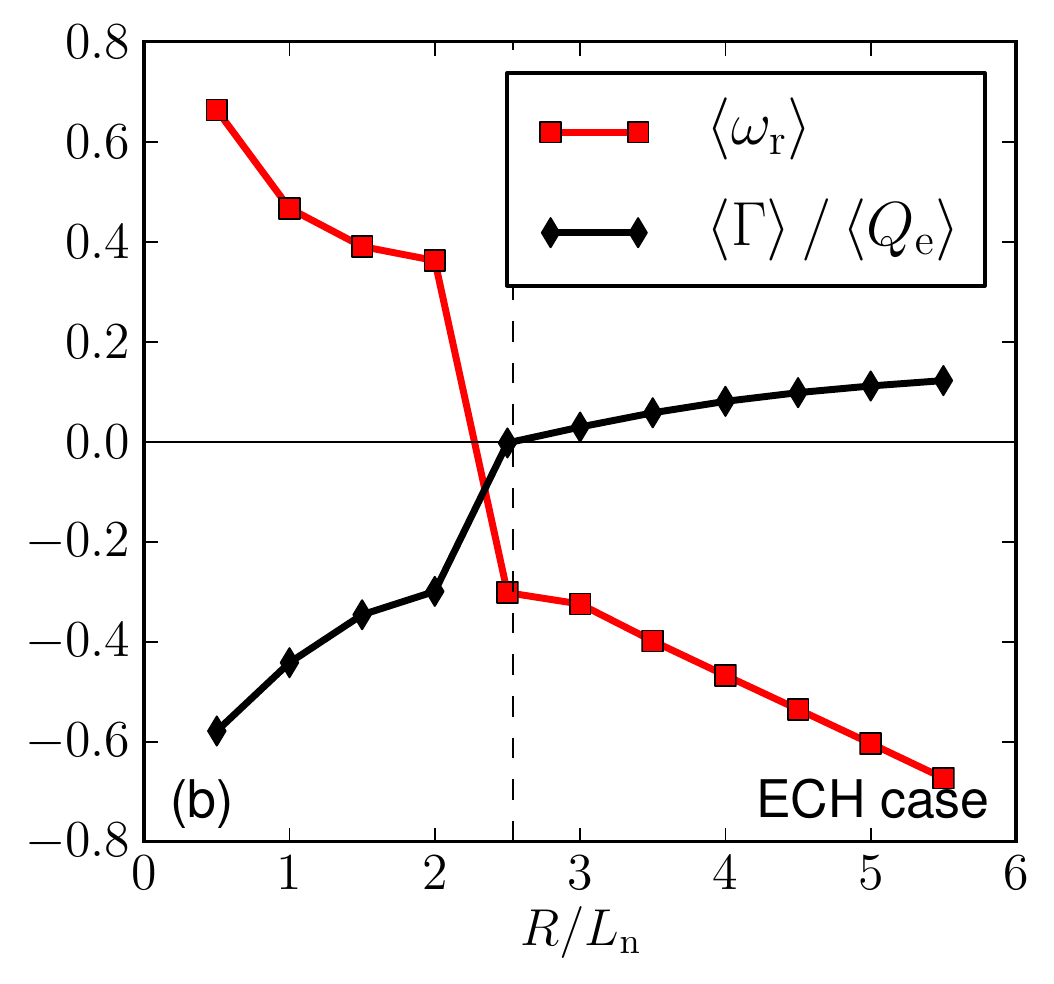}
    \caption{
    Quasi linear particle flux $\GammaQL$ (diamonds) and average mode frequency
    $\omegaQL$ as a function of $\ilrln$ \panel{a} OH reference case
    (\OHreference), \panel{b} ECH reference case (\ECHreference).
    }
    \label{fig:fprime-scan}
\end{myfigure}

In the absence of core particle sources, as it is the case in the TCV OH/ECH
H-modes \cite{zabolotsky2006influence}, the stationary value of the density
gradient $\ilrlnstat$  corresponds to the $\Gamma=0$ condition.  Here we
only consider the particle flux ascribed to turbulence, we shall deal with
neoclassical contributions later, in section \ref{sec:ware-pinch}.  Other
contributions to the total particle flux (e.g.~MHD, ripple losses, etc)
are neglected.  It is also assumed that these approximations hold under both
OH and ECH conditions. 
$\ilrlnstat$ is obtained by making a set of simulations for different values
for $\ilrln$, keeping all the other input parameters fixed.  At each
$\ilrln$, $\GammaQL$ is evaluated with equation \ref{eq:qlrule}.  Then the zero
point is found by linear interpolation.  

In order to understand the relationship between the linear and quasi-linear
fluxes, in figure \ref{fig:spectra-examples} we show the different contributions
in equation \ref{eq:qlrule} together with the mode frequency and growth rate of the
most unstable mode as a function $\aky$ for the OH reference case.  First we
note that the quasi-linear rule gives small weight at large wavenumbers (short
scales), therefore $\GammaQL$ and $\omegaQL$ are mainly determined by the
$\aky<\ $0.4--0.5 wavenumber range (ITG and TEM range).

When the density gradient $\ilrln$ is small (figure \ref{fig:spectra-examples} a),
b)), the most unstable modes are ITGs (positive $\omegaR$), the $w_k \Gamma^k$
contributions to $\GammaQL$ throughout the covered wavenumber range is negative
(inwards), making $\GammaQL$ pointing unambiguously inwards.  Increasing
$\ilrln$ (figure \ref{fig:spectra-examples} c) and d)), the growth rate of the most
unstable modes decreases and at low $\aky$, $w_k \Gamma^k>0$ contributions
appear, making $\GammaQL=0$ when $\ilrln=\ilrlnstat$ is reached.  Further
increasing $\ilrln$ ((figure \ref{fig:spectra-examples} e) and f)) the mode frequency
changes sign and TEM modes appear at small wavenumbers.  When these modes
dominate, the particle flux $\GammaQL$ is positive (outwards).

figure \ref{fig:fprime-scan} shows the quasi-linear particle flux $\GammaQL$
normalized to the total electron heat flux as a function of $\ilrln$ for the
two reference cases.  Each point in the figure is a result of 13 simulations
(averaged over the $k_y$ spectrum as in figure \ref{fig:spectra-examples}).
The average mode frequency $\omegaQL$ is also plotted showing that, indeed,
the zero flux point is found between ITG and TEM type modes
\cite{fable2010role}.  Note the role of ITG-TEM balance at a certain density
gradient in the spectrum and also in adjusting the value of $\ilrlnstat$.
It must also be stressed out, that at the co-existence of different modes,
one must certainly account for non-linear effects \cite{ernst2004role,
ernst2006identification, lapillonne2011nonlinear, gorler2011flux}.  It is
not expected that the absolute value of $\ilrlnstat$ from the quasi-linear
estimate of the particle flux instantly matches the experimental values;
however, its parametric dependence on the plasma parameters is remarkably
well retained \cite{fable2010role, fable2008eITB, angioni2011hmodes}.

\subsection{\texorpdfstring{$\nueff$--$\ilTeTi$ scan}{nueff - Te/Ti scan}}
\label{ssec:nueff-teti}

We endeavour to reproduce the collisionality dependence reported in
Sec.~\ref{ssec:collisionality-dependence} with numerical simulations.  Since
in TCV H-mode experiments collisionality is mainly controlled by intense
electron heating, the $\ilTeTi$ ratio certainly increases with additional
ECH.  It is expected that the temperature ratio has a significant effect on
the nature of the dominant instabilities and hence on the particle flux.  We
also want to explore the effect of the electron-ion collisions at each
considered temperature ratio. Therefore, we perform a double parameter scan
in the range of $\vnewk=[0.008, 0.012, 0.016, 0.020, 0.024]$ together with a
scan in $\ilTeTi=[1, 1.5, 2]$.  

The results are summarized in figure \ref{fig:rln_vs_nueff} a), which shows
the predicted value of the density gradient $\ilrlnstat$ as a function of
the collisionality $\vnewk$.  Each point is obtained from a scan over \ilrln
to determine \ilrlnstat where $\GammaQL=0$.  Therefore each point in
figure \ref{fig:rln_vs_nueff} is the result of about $10 \times 13$ linear
simulations.  When $\ilTeTi=1.0$ the predicted density gradient is
decreasing with collisionality.  As the temperature ratio is increased, this
trend is reversed and at $\ilTeTi=2.0$ an increase of $\ilrlnstat$ is
predicted with increasing $\vnewk$.  These results show a clearer trend
(figure \ref{fig:rln_vs_nueff} b)) if now one plots $\ilrlnstat$ as a function
of $\omegaQL$ which characterizes the nature of the background turbulence
and orders very well the predicted values of $\ilrlnstat$
\cite{fable2010role, angioni2011hmodes} and also clearly shows that the
change in the dominant instabilities alters the collisionality dependence.
Therefore, an explanation of the different behaviour of TCV H-modes with
respect to JET/AUG plasmas can be due to the different $\ilTeTi$ ratio and
hence a different collisionality dependence.

Collisionality provides a purely convective term to the particle flux
\cite{angioni2009particle_transport} which is directed outwards for ITG and
inwards for TEM modes and increases with increasing collisionality.
Intuitively one expects that increasing $\vnewk$ tends to push $\omegaR$
towards more positive values \cite{angioni2009particle_transport} (towards
ITG type turbulence) due the stabilizing effect of collisions on TEM
turbulence \cite{ernst2004role, angioni2005collisionality}.  The net effect
of collisionality on $\ilrlnstat$ stems from these two effects: $\vnewk$
influences the turbulence regime and, depending on the turbulence regime,
the particle flux.  Since an increase in collisionality implies an increase
in the real mode frequency the effect of the collisionality on the TEM modes
driven outward flux is expected to be weak
\cite{angioni2009particle_transport}.  Indeed, in
figure \ref{fig:rln_vs_nueff} we observe almost no collisionality dependence
at $\ilTeTi=1.5$, where $\omegaQL\approx0$.  Moving away from this point
toward either more positive or more negative $\omegaQL$, the net particle
flux is less inwards, and \ilrlnstat is found at a lower gradient.  The
increase of $\ilTeTi$ at fixed $\Ti$ is destabilizing for ITG and enhances
the TEM \cite{lang2007gyrokinetic} activity as well.  Their interplay is
such that TEM takes over rapidly with increasing $\ilTeTi$, in particular
with ECH i.e.~by increasing $\Te$ \cite{bottino2006linear}.  Indeed,
starting from the OH reference point in the ITG instability domain, with
increasing $\ilTeTi$ and decreasing $\vnewk$, $\omegaQL$ decreases and the
predicted value of $\ilrlnstat$ is increasing.  In the transition region
between ITG and TEM, the largest $\ilrlnstat$ is obtained, then moving
towards the TEM domain the density peaking decreases, leaving the point
corresponding to the ECH reference case with a very similar $\ilrlnstat$
value.

We conclude that albeit ECH has a significant effect on the values of $\vnewk$
and $\ilTeTi$, the change in these parameters only is not sufficient to explain
the observed overall experimental behaviour.

\begin{myfigure}
    \includegraphics[width=\imagewidth]{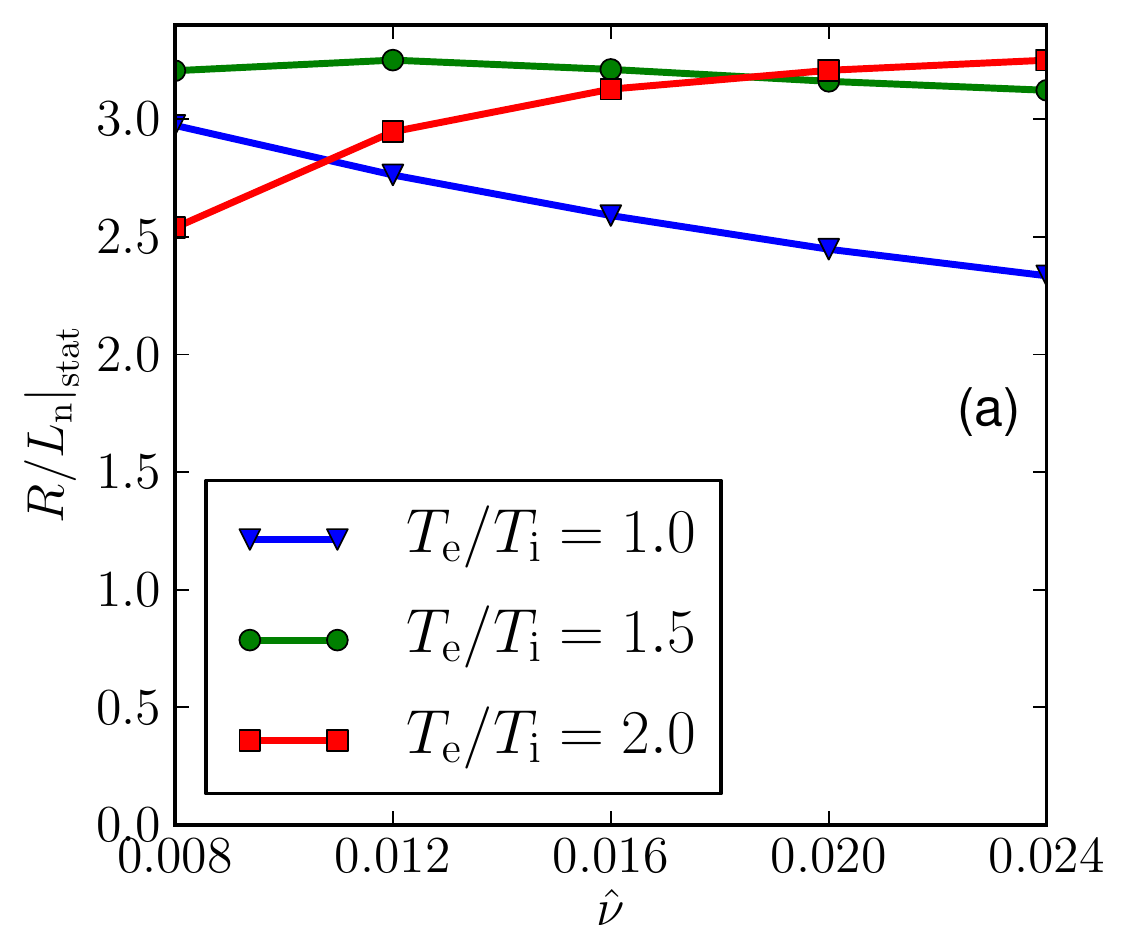}
    \includegraphics[width=\imagewidth]{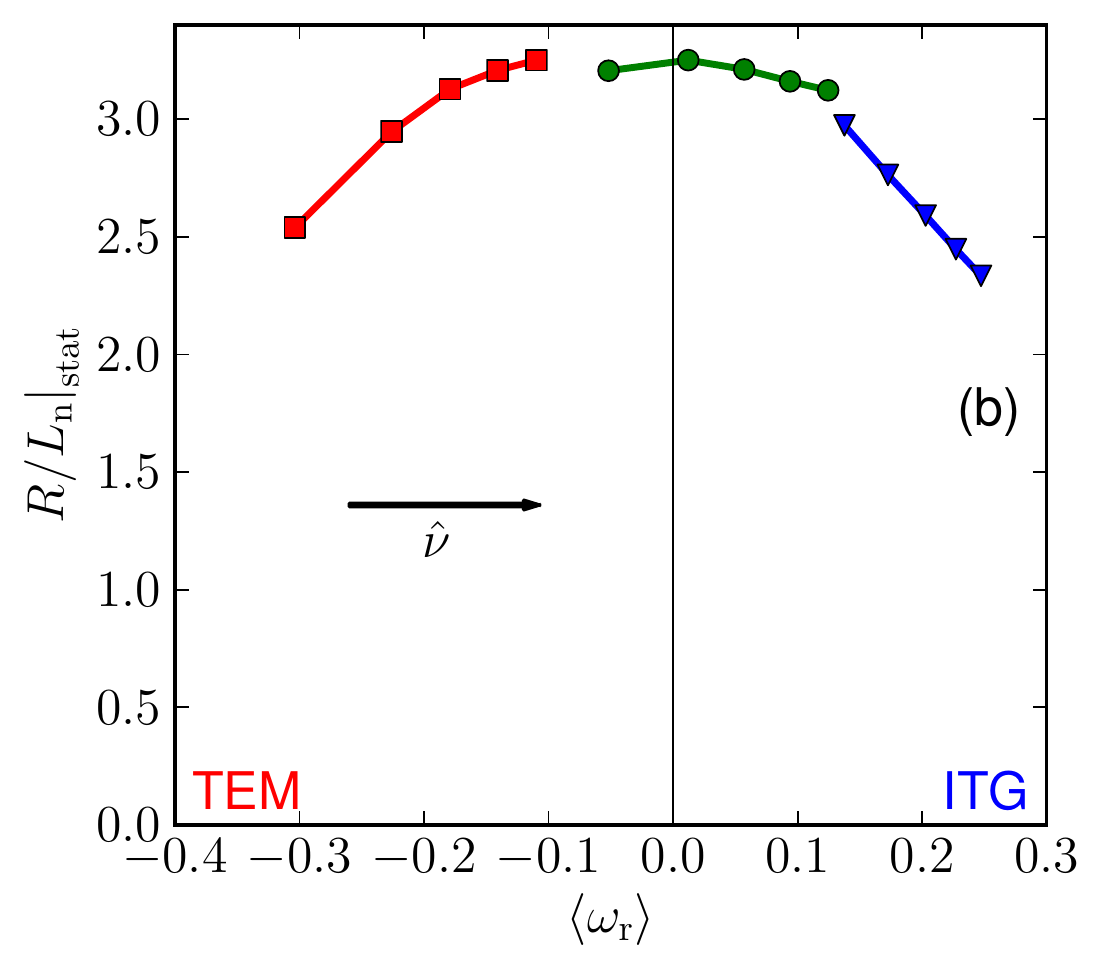}
    \caption{
    Predicted values of $\ilrln$ for different values of the
    temperature ratio $\ilTeTi$, as a function of \panel{a} the collisionality
    $\vnewk$ and \panel{b} the average mode frequency $\omegaQL$.
    }
    \label{fig:rln_vs_nueff}
\end{myfigure}

\subsection{\texorpdfstring{$\ilrlTe$ and $\ilrlTi$ dependence}{R/LTe and R/LTi
dependence}}
\label{ssec:rlt}

It is expected that the temperature gradients have a large effect on the
predicted value of the density gradient.  As it was pointed out in
Sec.~\ref{ssec:ion_measurements}, the ion temperature gradient is generally
lower than that of the electrons even at the high collisionality OH plasmas.
The ions are even more decoupled from the electrons when intense electron
heating is applied.  Thus we expect that $\ilrlTi$ does not change
significantly in these plasmas (more precisely it does not increase).  The
experimental data are more clear on the $\ilrlTe$ dependence; however it is
useful to study the dependencies on the parameter, as well.

Two scans of $\ilrlTe$ in the range of $[6, 7.5, 9, 10.5, 12]$ have been
performed: the first with OH only parameter set (\OHreference); the second with
ECH parameter set (\ECHreference).  The other parameters were kept fixed
(c.~f.~Sec.~\ref{ssec:simulation-setup}).  In the same way, we change the value
of $\ilrlTi$ in the range of $[4, 5, 6, 7.5, 9]$.

In figure \ref{fig:rln_vs_tprim} we present the results of these four scans in
terms of $\eta_\mathrm{ei} = \left(\ilrlTe\right)/\left(\ilrlTi\right)$.
Figure \ref{fig:rln_vs_tprim} a) shows that the change in $\ilrlTi$ has a
different effect on the OH and the ECH reference cases.  Indeed, starting with
the OH reference case (triangles), increasing $\ilrlTi$ predicts smaller
$\ilrln$. On the other hand, this parameter has the opposite effect on the
predicted peaking for the ECH case.  In figure \ref{fig:rln_vs_tprim} b), it is
shown that increasing $\ilrlTe$ increases $\ilrln$ for the OH case, while the
density peaking is rather insensitive to the change in the electron temperature
gradient for the ECH case.

In figure \ref{fig:ctcp_vs_tprim} we decompose the predicted $\ilrln$ in terms of
$\CT$ and $\CP$ as in equation (\ref{eq:rlnstat}) in order to identify the change in
the thermodiffusive and the other contributions.  We see that the larger
contribution is always the thermodiffusive pinch $\CT \ilrlTe$ and usually the
two pinches change in the opposite direction and a complicated interplay between
the two yields the predicted $\ilrln$.  This means that the parametric
dependence of the density peaking is largely determined by the starting
reference parameters as mentioned above.

In figure \ref{fig:rln_ctcp_vs_omega} a) we plot these results against $\omegaQL$
which, again, orders very well the simulation results $\ilrln$
\cite{fable2010role, angioni2011hmodes}.  As it was already shown in
Sec.~\ref{ssec:nueff-teti}, the OH reference (shaded triangle) case is in the
ITG regime ($\omegaQL > 0$), while the ECH (shaded square) reference case is
predicted to be more TEM ($\omegaQL < 0$).  Increasing the electron temperature
gradient results in moving to the left on the $\omegaQL$ axis (towards negative
values), while the ion temperature gradient pushes towards larger $\omegaQL$
values.  The density peaking is maximal around $\omegaQL \approx 0$ and the
predicted value decreases towards both ITG and TEM regime as predicted for eITB
and L-mode parameters \cite{fable2008eITB, fable2010role}. We therefore show
that this is a very general feature.  In figure \ref{fig:rln_ctcp_vs_omega} b) c)
we see the thermodiffusive contribution increases towards the ITG regime, while
the $\CP$ term decreases.  This behaviour of the two pinch terms reinforces
their suggested universal role in particle transport \cite{fable2010role}.

As mentioned in section \ref{ssec:ion_measurements}, the ion temperature
measurements indicate that $(\ilrlTi)/(\ilrlTe) \approx 0.5$, although we do
not have measurements near $\rho_\psi=0.7$, where we have based our
analysis.  If this ratio is further decreased in the simulations, moving
left in figure \ref{fig:rln_ctcp_vs_omega} a), the value of $\ilrlnstat$
decreases, following the trend seen previously.  However the range of $\aky$
where the ITG is the most unstable mode, typically $\aky=0.1-0.6$, shrinks
and slab-like electronic modes very elongated along the magnetic field lines
appear as seen in Refs.~\cite{hallatschek2005giant, angioni2007non}.
Although their contribution to the particle flux is small, so is that of the
remaining ITG and TEM.  This domain requires non-linear simulations and the
role of non-adiabatic passing electrons to be studied in detail, which is
left for further analyses.

\begin{myfigure}
    \includegraphics[width=\imagewidth]{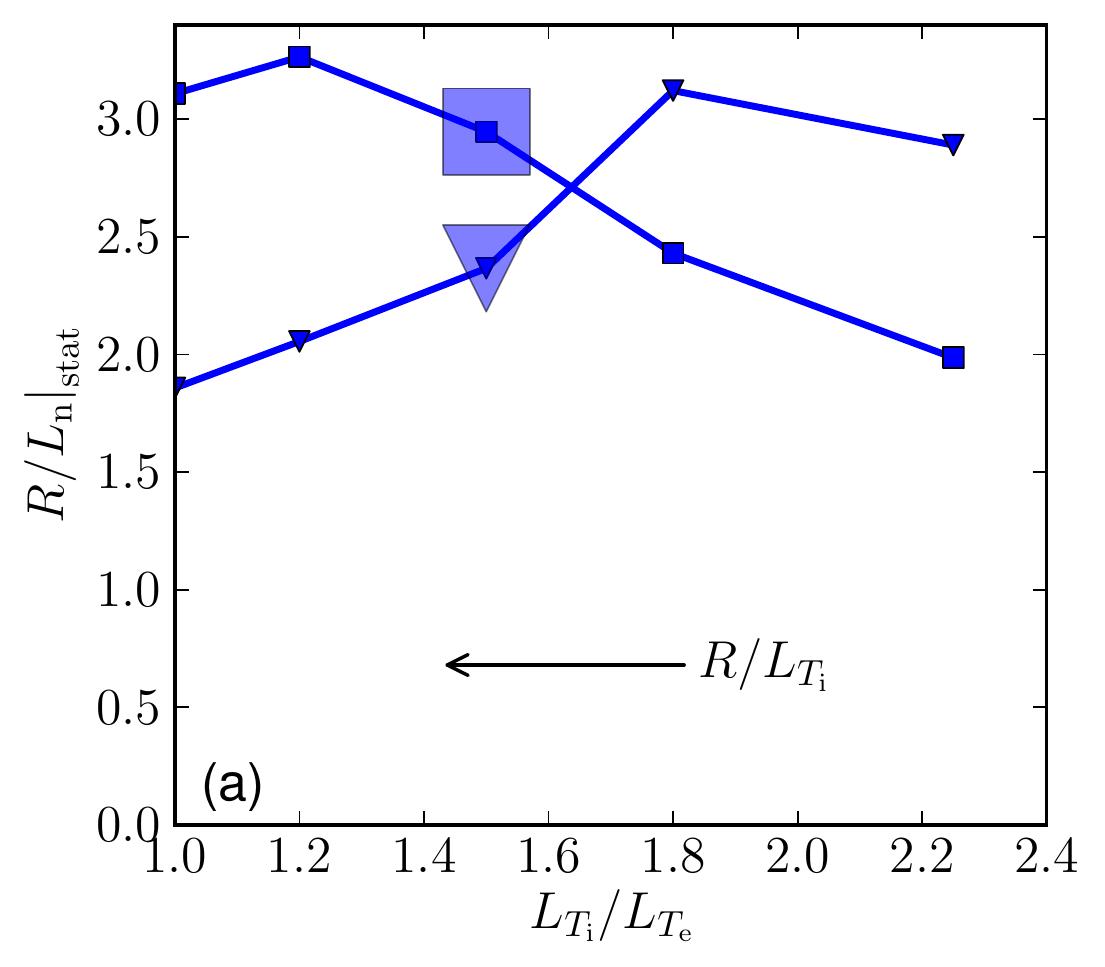}
    \includegraphics[width=\imagewidth]{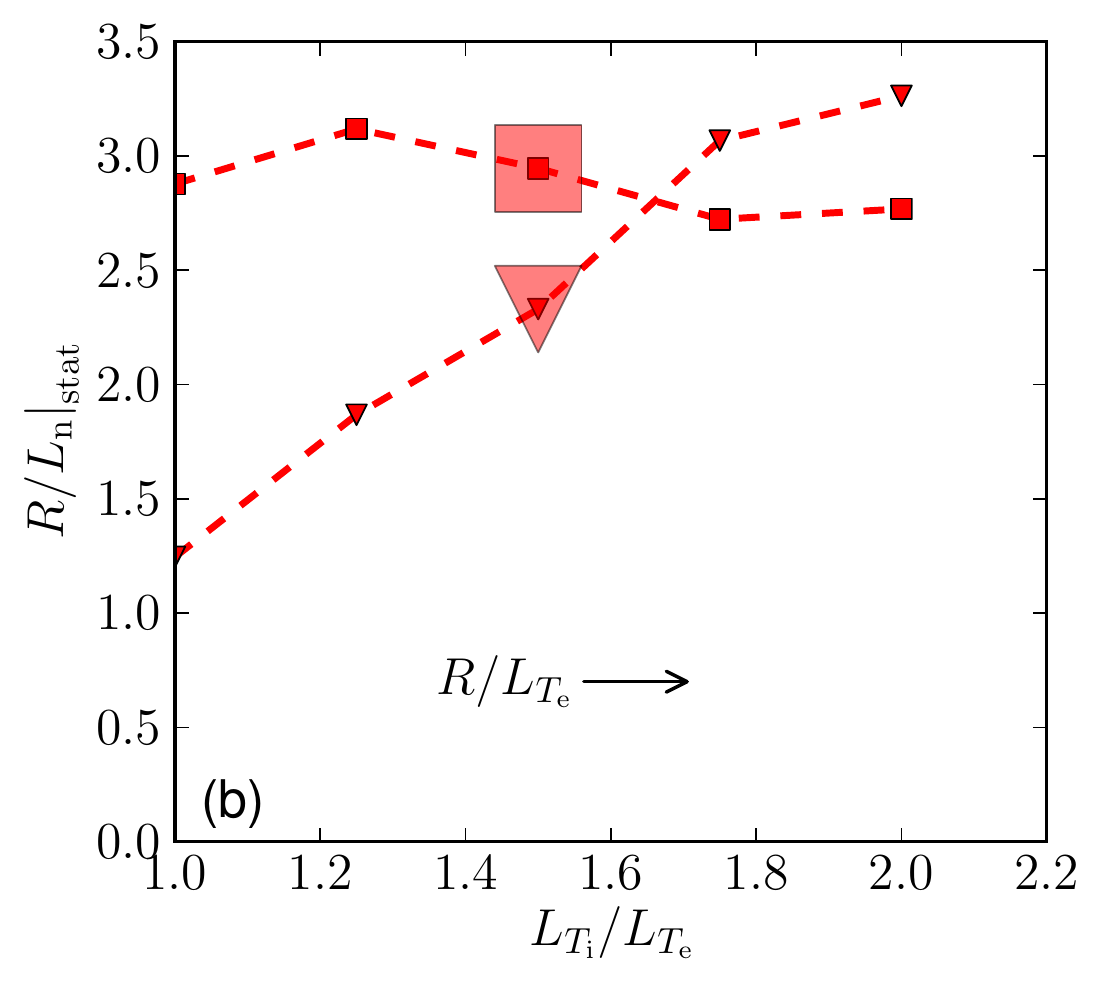}
    \caption{
    Predicted values of normalized density gradient $\ilrln$ as a function of
    the temperature gradient ratio $\ilgradTrat$ changing $\ilrlTi$ (solid) and
    $\ilrlTe$ (dashed).  Different symbols indicate different heating phases,
    with OH only (triangles) and ECH (squares).  The larger shaded symbols show
    the reference cases.
    }
    \label{fig:rln_vs_tprim}
\end{myfigure}

\begin{myfigure}
    \includegraphics[width=\imagewidth]{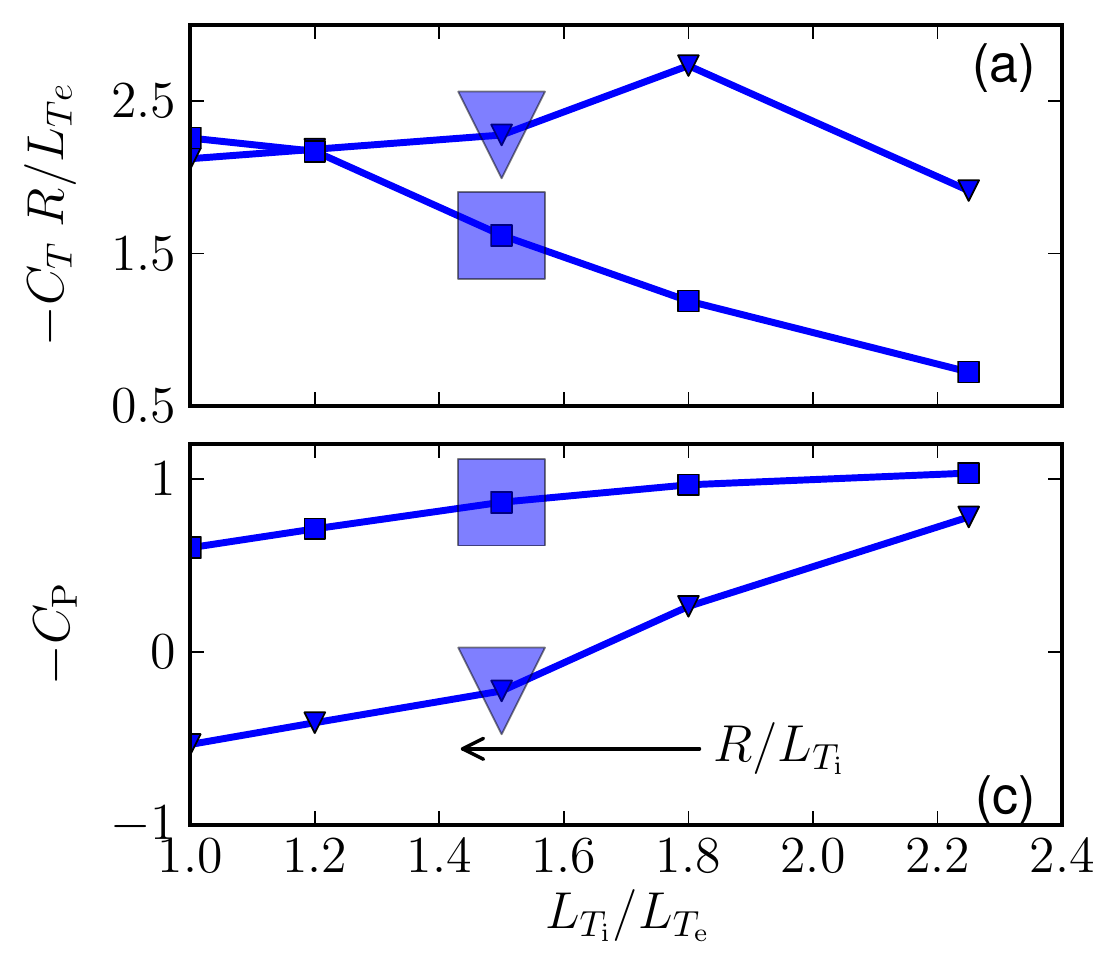}
    \includegraphics[width=\imagewidth]{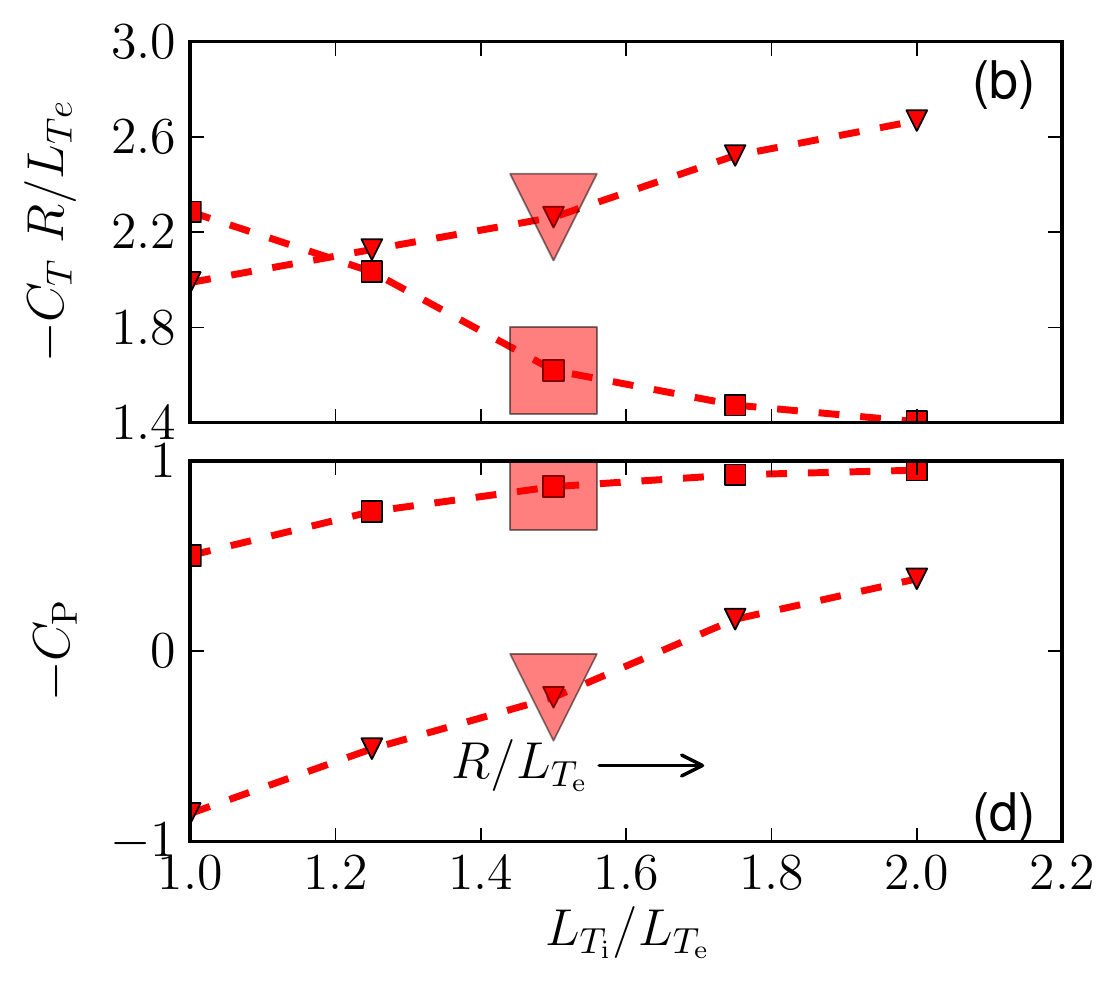}
    \caption{
    Thermodiffusive $-\CT \ilrlTe$ and the other pinch $-\CP$ contributions to
    the predicted value of $\ilrln$ (c.~f.~ figure \ref{fig:rln_vs_tprim}) as a
    function of the temperature gradient ratio $\ilgradTrat$ changing $\ilrlTi$
    (solid) and $\ilrlTe$ (dashed).  Different symbols indicate different
    heating phases, with OH only (triangles) and ECH (squares).  The larger
    shaded symbols show the reference cases.
    }
    \label{fig:ctcp_vs_tprim}
\end{myfigure}

\begin{myfigure}
    \includegraphics[width=\imagewidth]{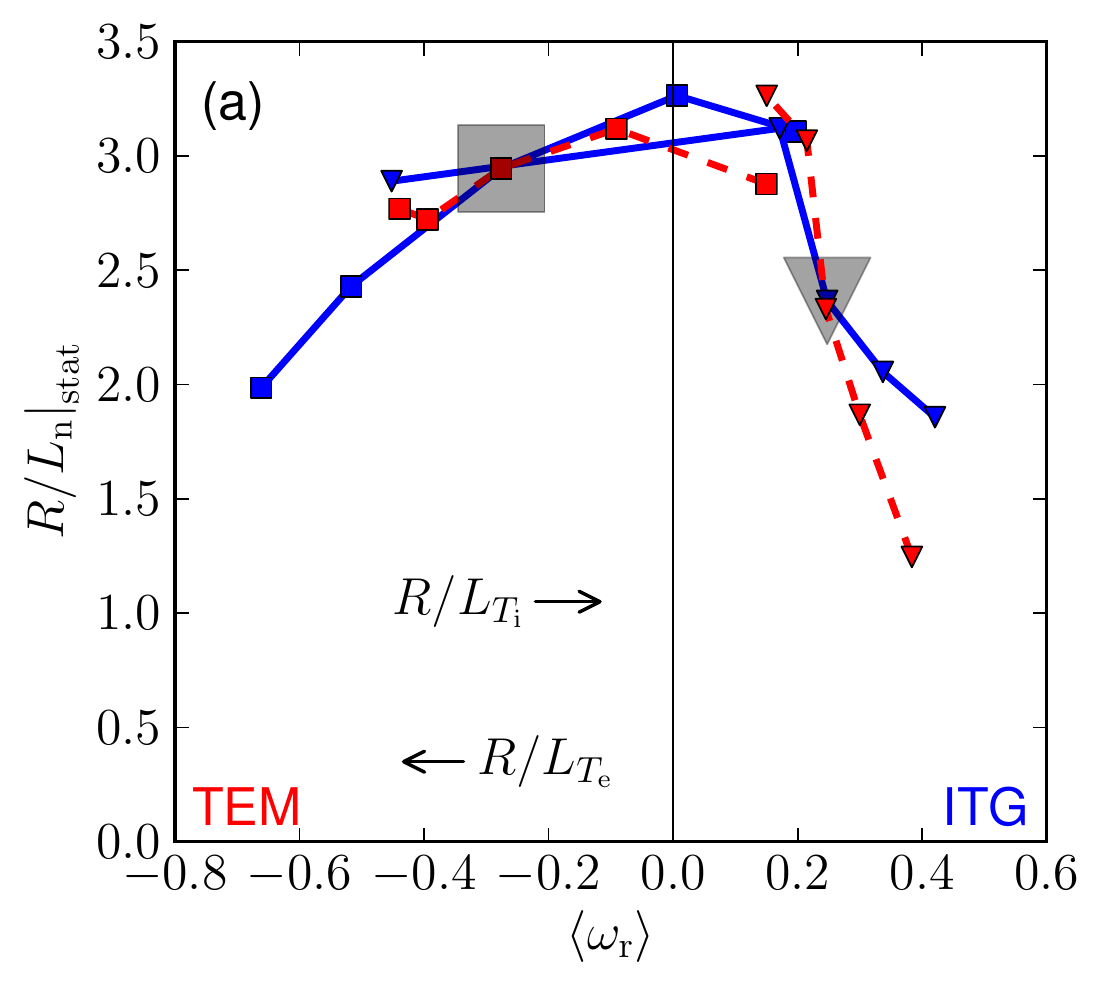}
    \includegraphics[width=\imagewidth]{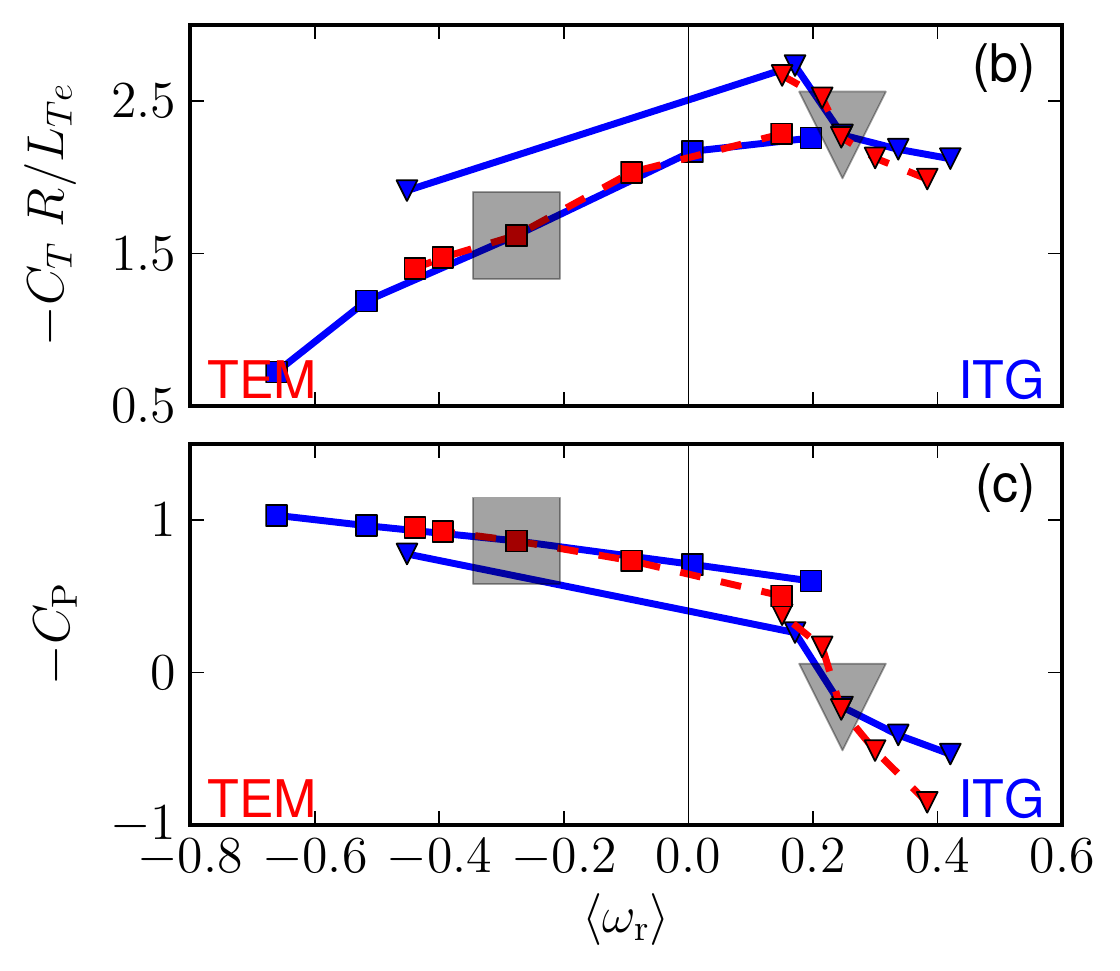}
    \caption{
    The predicted value of $\ilrln$, the  thermodiffusive $-\CT \ilrlTe$ and the
    other pinch $-\CP$ contributions as a
    function of the average mode frequency $\omegaQL$ changing $\ilrlTi$
    (solid) and $\ilrlTe$ (dashed).  Different symbols indicate different
    heating phases, with OH only (triangles) and ECH (squares).  The larger
    shaded symbols show the reference cases.
    }
    \label{fig:rln_ctcp_vs_omega}
\end{myfigure}

\subsection{Effect of the Ware-pinch}
\label{sec:ware-pinch}

In ohmic plasmas, the neoclassical Ware-pinch can have significant contribution
to the density peaking \cite{ernst2004role, stober2003dependence}.  The strength
of this effect depends also on the heating scheme applied \cite{ernst2004role,
angioni2011hmodes}.  In the presence of the Ware-pinch, one needs to find
$\ilrlnstat$ from:
\( \Gamma + \dense \Wp=0  \)
\cite{ernst2004role}.
In order to cancel normalization factors, it is more convenient to use the
$\Gamma/\qe$ ratio; therefore, we divide with the electron heat flux $\qe$:
\begin{equation}
    \frac{\Gamma}{\qe}+ \frac{\dense \Wp}{\qe}=0.
\end{equation}
We evaluate the first term using the ratio of the normalized fluxes
$\GammaQL/\qeQL$ (which is proportional to $\Te \Gamma/ \qe$)
from numerical simulations, while the second term is computed from experimental
measurements using $\qe = \qeEXP$ and power balance (PB) arguments assuming a
stationary plasma, where
\begin{equation}
    \qeEXP = -\ChiePB \dense \nabla \Te = \ChiePB \rlTe
    \frac{\dense \Te}{R}.
    \label{eq:Q_exp}
\end{equation}
We find
\begin{equation}
    \frac{\GammaQL}{\qeQL}+ \frac{ R \Wp}{\ChiePB \ilrlTe} = 0,
    \label{eq:pflux_ware}
\end{equation}
from which one can find the stationary density gradient value with the
Ware-pinch effect included. Note that $\Wp < 0$.  Note also that for any
consistent normalization of $\GammaQL$ and $\qeQL$, $\Te$ of
equation \ref{eq:Q_exp} cancels in equation \ref{eq:pflux_ware}.

We compare the Ware-pinch contribution for the OH and ECH reference cases
substituting typical values of $\Wp$ and $\ChiePB$ in
equation \ref{eq:pflux_ware}, assuming that the intense ECH does not directly
drive significant particle flux that must be countered by the turbulent flux
to restore the total flux to zero.
In figure \ref{fig:ware-effect} the ratio of the normalized fluxes
$\GammaQL/\qeQL$ without (symbols), and with (no symbols) the Ware-pinch
contribution are shown as a function of $\ilrln$; (a) shows the OH reference
case with $\ChiePB=0.3\,\mathrm{m^2 s^{-1}}$, (b) is the ECH reference case
with $\ChiePB=1\,\mathrm{m^2 s^{-1}}$. For both cases $\Wp =
-0.3\,\mathrm{ms^{-1}}$.
It can be seen that the shape of the $\GammaQL$ curve around $\ilrlnstat$
can influence the position of the new value of $\ilrlnstat$.  The addition
of the Ware-pinch results in a vertical downshift of the curve.  At the OH
reference case $\GammaQL$ is almost linear with $\ilrln$.  Due to its modest
slope, the upshift in \ilrlnstat can be as large as 1.  For the ECH case the
slope of the $\Gamma$ curve is somewhat reduced which would imply a larger
change in $\ilrlnstat$.  However, for the ECH case the contribution of the
Ware-pinch is negligible mainly because of two reasons.  First, the loop
voltage decreases as the electron temperature increases at constant current
\cite{ernst2004role}.  This reduces the absolute value of $\Wp$.  Second,
due to the profile stiffness the value of $\ChiePB \ilrlTe$ is larger in ECH
plasmas, which makes the contribution of the second term in
equation \ref{eq:pflux_ware} much smaller.

\begin{myfigure}
    \includegraphics[width=\imagewidth]{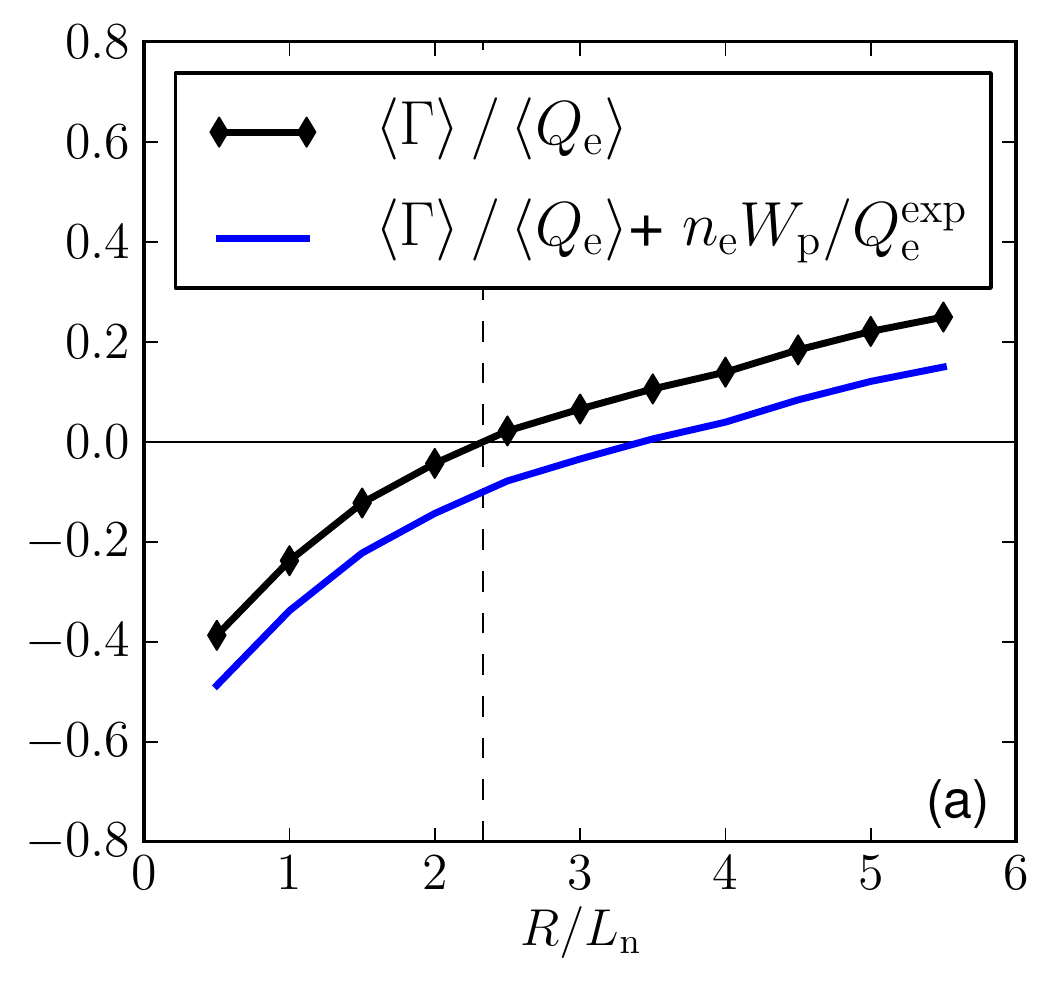}
    \includegraphics[width=\imagewidth]{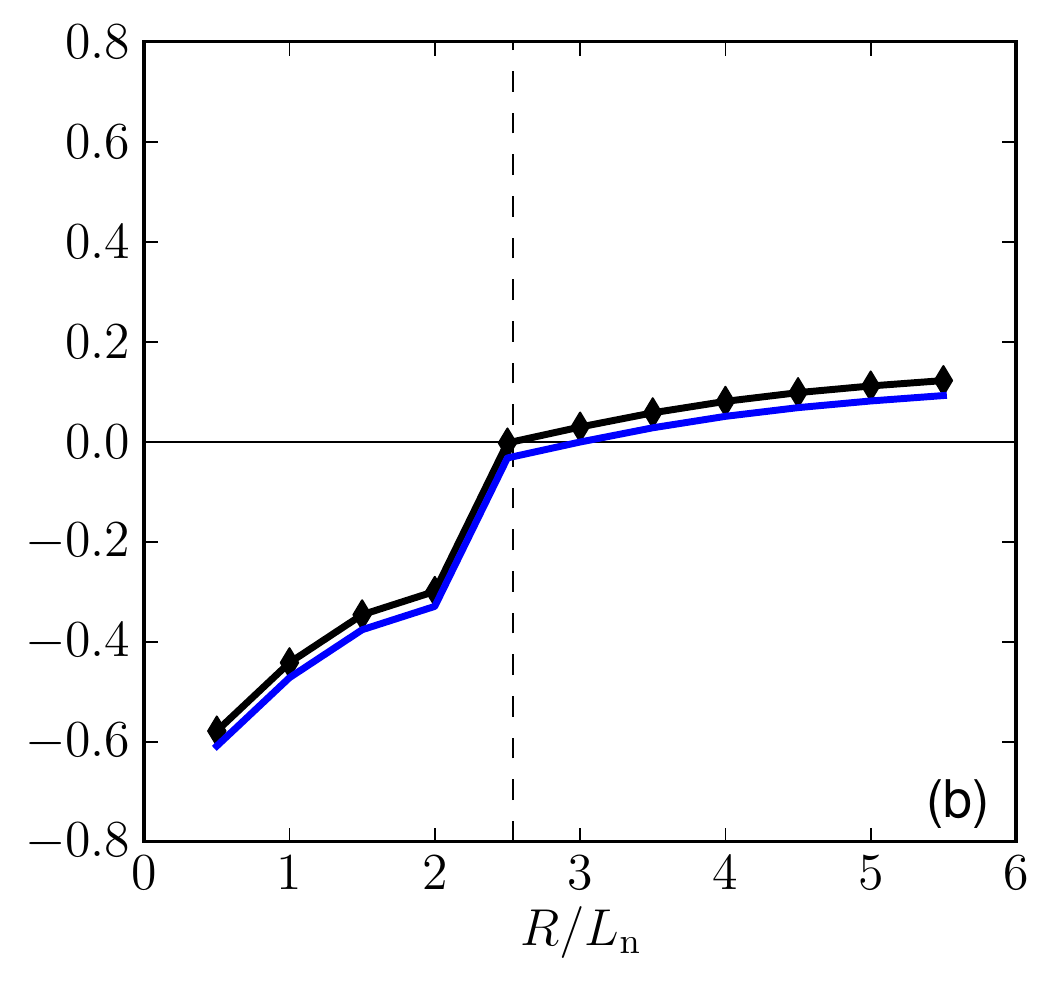}
    \caption{Normalized particle flux as a function of $\ilrln$, with (no
    symbols) and without (diamonds) the Ware-pinch contribution
    according to equation (\ref{eq:pflux_ware}):
    (a) OH model scenario, $\ChiePB=0.3\ \mathrm{m^2/s}$
    (b) ECH model scenario, $\ChiePB=1\ \mathrm{m^2/s}$.
    For both cases: $\Wp=-0.3\ \mathrm{m/s}$. }
    \label{fig:ware-effect}
\end{myfigure}

\subsection{Choice of the quasi-linear rule}
\label{sec:qlrule}

Our analysis is consistent with the observed experimental data.  However it
is based on a specific quasi-linear rule (equation \ref{eq:power_weights} with
$\xi = 2$).  Other similar studies reported successful interpretations with
different quasi-linear theory \cite{angioni2011hmodes, bourdelle2007new,
casati2009validating, maslov2009density, merz2010nonlinear, fable2008eITB}
using different summation rule, or even using only one $\ky$ mode.  A
question arises whether the results reported here are strongly dependent on
the choice of the $w$ weights in equation \ref{eq:qlrule}.

We evaluated the quasi-linear fluxes and $\ilrlnstat$ with several different
choices of $w$ for the parameter scans discussed previously.  In
figure \ref{fig:many-ql-rules} we show again the results of the temperature
gradient scans (see figure \ref{fig:rln_ctcp_vs_omega}) summing many modes
with the power law used thorough this paper and Ref.~\cite{fable2010role}
(squares), summing with the classical mixing-length estimate (diamonds),
summing with the exponential rule (triangles) motivated by fluctuation
measurements \cite{bourdelle2007new}, using one mode where the mixing length
transport is maximal (circles) \cite{fable2008eITB} and using one mode with
$\aky = 0.3$ (stars) \cite{angioni2011hmodes}.
Although, the differences in the predicted absolute value of $\ilrlnstat$
can be as large as $1.5$, the relative changes in $\ilrlnstat$ and the
parametric dependencies are nearly identical, independent on the choice of
the quasi-linear rule.
Figure \ref{fig:many-ql-rules} and \ref{fig:rln_ctcp_vs_omega} show that the
main trends do not depend so much on the quasi-linear rule but rather on the
main dependence of the underlying most unstable modes.  However, it is safer
to use the whole spectrum in order to be sure to capture both the effects of
ITG and TEM modes.  The importance of the most unstable modes explains why
similar trends can be expected in non-linear simulations
\cite{merz2010nonlinear, lapillonne2011nonlinear, gorler2011flux} and in the
experiments.

\begin{myfigure}
    \includegraphics[width=\imagewidth]{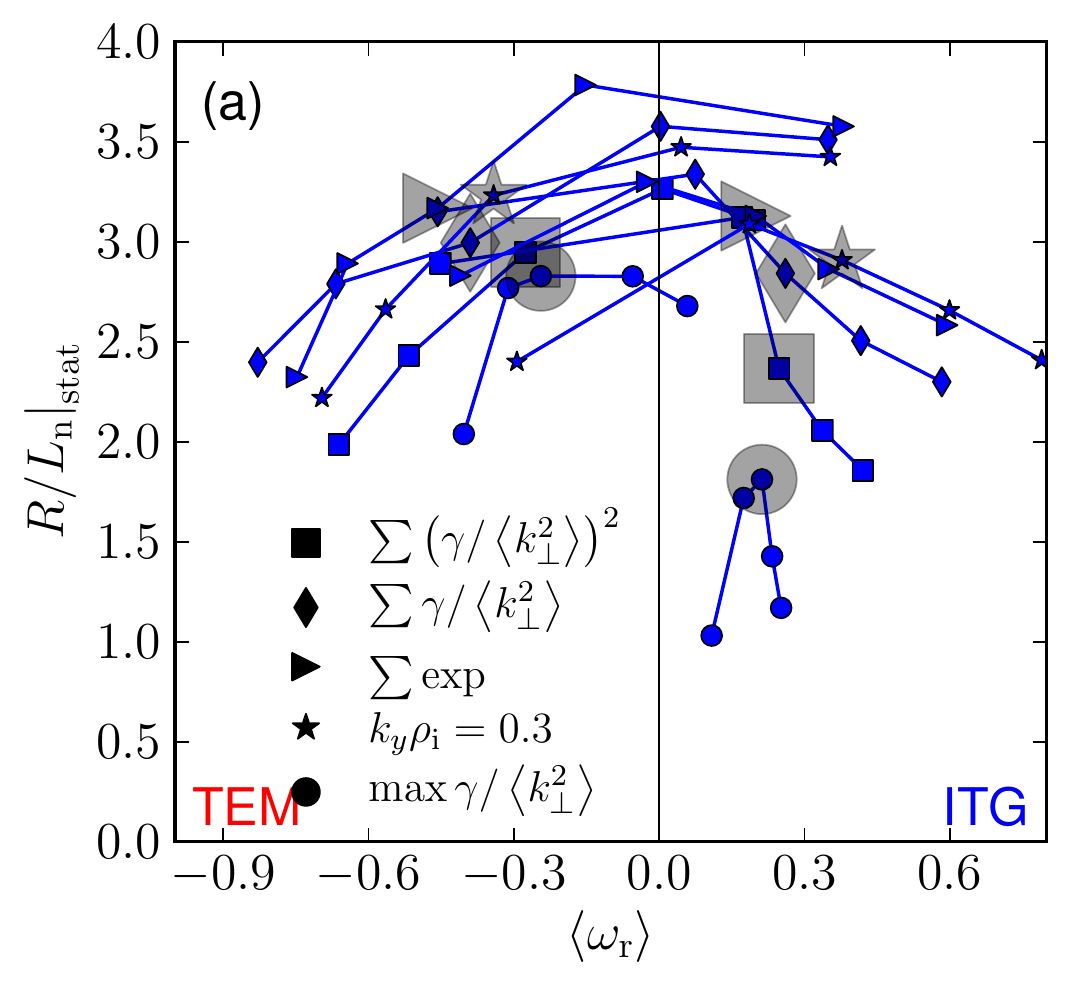}
    \includegraphics[width=\imagewidth]{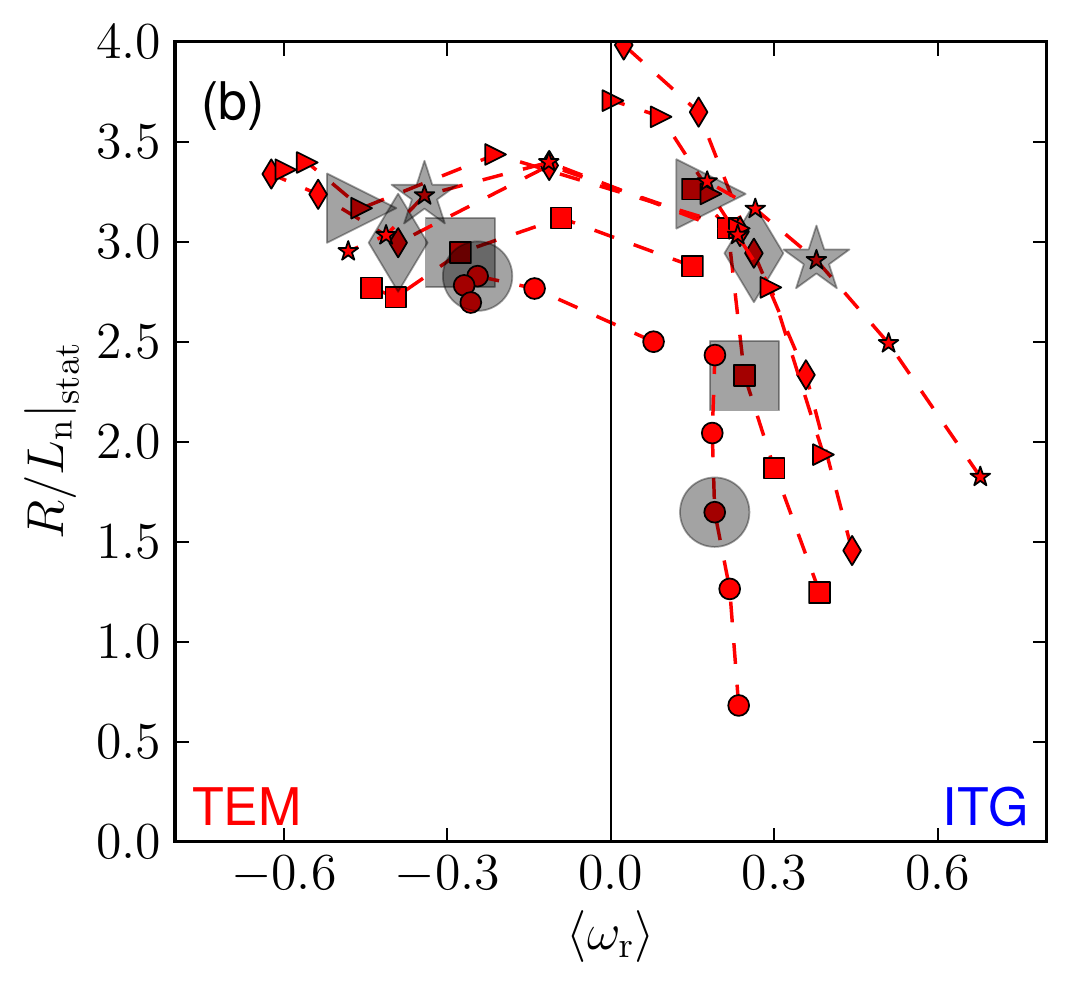}
    \caption{%
    Same as figure \ref{fig:rln_ctcp_vs_omega} \panel{a}, using different
    quasi-linear rules: \panel{a} $\ilrlTi$ scan, \panel{b} $\ilrlTe$ scan.
    Different symbols show different prescriptions for the $w$ weights in
    equation \ref{eq:qlrule}.  The larger shaded symbols mark the reference cases
    as in figure \ref{fig:rln_ctcp_vs_omega}: the OH case for $\omegaQL>0$ and
    ECH case otherwise.
    }
    \label{fig:many-ql-rules}
\end{myfigure}

\section{Conclusions}
\label{sec:discussion}

The collisionality dependence of the local logarithmic electron density
gradient $\ilrln$ has been investigated in ohmic and electron cyclotron
resonance heated H-mode plasmas on TCV.  The density profile flattens when
ECH power is added, the normalized density gradient $\ilrlnstat$ at
$\rho_\psi\approx0.7$ decreases from about 3 to approximately 1.5.  The
dependence of $\ilrlnstat$ on collisionality is rather unusual: the density
peaking increases with increasing collisionality.  With local quasi-linear
gyrokinetic simulations it was shown that the change in collisionality alone
cannot reproduce the observed experimental behaviour.  The simulation
results show that the most important parameter is the ion temperature
gradient $\ilrlTi$, which was not available in these experiments.  Using
$\ilrlTi = 6$ puts the OH and the ECH reference case in the ITG and the TEM
regime, respectively, with predicted $\ilrln \approx 2.5$ and $3$.  For the
OH case the contribution of the Ware-pinch can be as large as $\Delta \ilrln
\approx 1$ resulting in a peaking around 3.5 which is in good agreement with
the experiments.  However, the flat ECH profiles are not recovered.  If a
lower ion temperature gradient $\ilrlTi=4$ is assumed, the predicted OH
density gradient increases to around 4 (with the Ware pinch included) and
that of the ECH case decreases to 2.  The experimental trends are,
therefore, qualitatively explained.  The edge ion temperature data suggest
that $\ilrlTi$ is about 3-5, however new experimental data are required to
compare with our simulation results.  We have also seen that at very low
$\ilrlTi$, below 3-4, electron modes with large $\aky$ are seen to play a
role, therefore new simulations are also required in this domain with
non-adiabatic passing electrons and non-linear model.  The quasi-linear
simulations predict that $\ilrlnstat$ for the OH and the ECH reference case
decreases as $\ilrlTi$ is decreased, however dedicated experiments are
needed to explore how the $\Ti$ profile changes between OH and ECH
phases.

The simulation results, where the parameters can be changed individually,
also show that the collisionality dependence can actually be due to a change
in $\ilTeTi$ and $(\ilrlTe/\ilrlTi)$ rather than the collisionality itself.
Table \ref{tab:summary} summarizes the change in $\ilrlnstat$ due to the
considered parameters.

The $\GammaQL=0$ stationary condition is set by the balance between ITG and
TEM modes as obtained in L-mode simulations \cite{fable2010role}.  Depending
on the turbulence regime, the dependence of $\ilrlnstat$ on the
collisionality, temperature ratios and temperature gradients can be very
different, but in all cases the average mode frequency $\omegaQL$ as a
figure of merit of the background turbulence clarifies the simulation
results.  This property of real mode frequency of the most unstable modes
was observed in the study of L- and H-mode AUG plasmas
\cite{angioni2005relationship, angioni2011hmodes}, L-mode and eITB TCV
plasmas \cite{fable2008eITB, fable2010role}, and now has been confirmed for
TCV H-mode parameters, as well.  The neoclassical Ware-pinch can also
contribute to the peaking in certain conditions typically when the
stationary point lies in an ITG dominated regime and when the electron heat
flux is not too large.

It is interesting to relate the present results to those of AUG reported
recently in \cite{angioni2011hmodes}.  The density peaking in those
considered H-mode plasmas increases with increasing additional ECH power,
mainly due to the increase in the ratio of the logarithmic electron
temperature gradient to the logarithmic ion temperature gradient and partly
also from the increase in the electron to ion temperature ratio.  Figure 7.
\panel{a} in \cite{angioni2011hmodes} can be directly compared to the curve
with stars in figure \ref{fig:many-ql-rules}.  In purely NBI heated AUG
H-mode plasmas ITG modes are the most unstable modes, positioned on the
right part of the plot with positive frequencies.  Adding ECH destabilizes
the TEMs and $\ilrlnstat$ increases as the mode frequency decreases.  It is
predicted that even larger amount of ECH would result in decreasing peaking
\cite{angioni2011hmodes}, however, this could not be validated by those
experimental results.  According to our interpretation, the TCV OH plasmas
are positioned very close to the ITG-TEM boundary and the additional ECH
moves $\ilrlnstat$ down on the left branch of the curve in figure
\ref{fig:many-ql-rules}.  Therefore our results combined with AUG
observations \cite{angioni2011hmodes} recover the full non-monotonic
$\omegaQL$ dependence.

In these TCV H-mode plasmas, the local behaviour of the density profile can
be well explained in a unified quasi-linear gyrokinetic theory of ITG and
TEM instabilities and turbulence.  This work confirms that the model used to
understand particle transport in L-mode \cite{fable2010role} and eITB
\cite{fable2008eITB} plasmas is also able to reproduce the main trends
observed in H-modes.  The results of a quasi-linear theory used in this work
can be further improved and validated by non-linear simulations and also
global codes are expected to predict more precise values of $\ilrlnstat$.
For a better test of this theory, systematic and precise ion temperature
profile measurements are desirable.  The systematic study of the dependence
of local density peaking on the plasma parameters such as collisionality,
electron to ion temperature ratio, electron and ion temperature gradient,
Ware-pinch show that the value of the plasma parameters from which these
scans are performed are important in order to understand their effects on
the stationary value of the density gradient.  This complex interdependence
is found to be greatly simplified when analysed in terms of the quasi-linear
real mode frequency of the most unstable modes.

\begin{table}
    \centering
    \begin{tabular*}{.5\textwidth}{@{}@{\extracolsep{\fill}}llll@{}}
    \toprule
    $\vnewk$    & $0.5$      & $\nearrow$ \\
    $\ilTeTi$   & $0.5$      & $\searrow$ \\
    $\eta$      & $1 - 1.5$  & ECH $\searrow$    & OH $\nearrow$ \\
    $\Wp$       & $0.5 - 1$  & $\nearrow$ \\
    \bottomrule
    \end{tabular*}
    \caption{Change in $\ilrln$ when the related parameter increases.}
    \label{tab:summary}
\end{table}

\section*{Acknowledgements}

The authors thank M.~Kotschenreuther and W.~Dorland for having made available
the GS2 code.  The GS2 simulations have been run on the PLEIADES2 cluster
EPFL, Lausanne.  This work has been supported in part by the Swiss National
Science Foundation.

\section*{References}
\bibliographystyle{unsrt}
\bibliography{bibliography}

\end{document}